\documentclass[12pt,preprint]{aastex}

\slugcomment{AJ in press}

\shorttitle{ExploreNEOs paper I: First results}
\shortauthors{Trilling et al.}

\newcommand{\pv}{$p_V$}

\begin{document}


\title{ExploreNEOs I: Description and first results
from the Warm Spitzer NEO Survey}


\author{D. E. Trilling\altaffilmark{1}}

\author{M. Mueller\altaffilmark{2}}

\author{J. L. Hora\altaffilmark{3}}

\author{A. W. Harris\altaffilmark{4}}

\author{B. Bhattacharya\altaffilmark{5}}

\author{W. F. Bottke\altaffilmark{6}}

\author{S. Chesley\altaffilmark{7}}

\author{M. Delbo\altaffilmark{2}}

\author{J. P. Emery\altaffilmark{8}}

\author{G. Fazio\altaffilmark{3}}

\author{A. Mainzer\altaffilmark{7}}

\author{B. Penprase\altaffilmark{9}}

\author{H. A. Smith\altaffilmark{3}}

\author{T. B. Spahr\altaffilmark{3}}

\author{J. A. Stansberry\altaffilmark{10}}

\and

\author{C. A. Thomas\altaffilmark{1}}


\altaffiltext{1}{Department of Physics and Astronomy,
Northern Arizona University, Flagstaff, AZ 86001;
\email{david.trilling@nau.edu}}
\altaffiltext{2}{Universit\'e de Nice Sophia
Antipolis, CNRS, Observatoire de la C\^ote d'Azur, BP 4229, 06304 Nice
Cedex 4, France}
\altaffiltext{3}{Harvard-Smithsonian Center for Astrophysics, 60 Garden St., MS-65, Cambridge, MA 02138}
\altaffiltext{4}{DLR Institute of Planetary Research, Rutherfordstrasse 2, 12489 Berlin, Germany}
\altaffiltext{5}{NASA Herschel Science Center, Caltech, M/S 100-22, 770 South Wilson Ave.
Pasadena, CA  91125  USA}
\altaffiltext{6}{Southwest Research Institute,  1050 Walnut St, Suite 300, Boulder, Colorado 80302, USA}
\altaffiltext{7}{Jet Propulsion Laboratory, California Institute of Technology, Pasadena, CA 91109, USA}
\altaffiltext{8}{Department of Earth and Planetary Sciences,
University of Tennessee,
1412 Circle Dr.,
Knoxville, TN 37996}
\altaffiltext{9}{Department of Physics and
Astronomy, Pomona College, 610 N. College Ave,
Claremont, CA 91711}
\altaffiltext{10}{Steward Observatory,
University of Arizona,
933 N. Cherry Ave,
Tucson AZ 85721}


\begin{abstract}
We have begun the ExploreNEOs project in which we observe
some 700~Near Earth Objects (NEOs) 
at 3.6~and 4.5~microns
with
the Spitzer Space Telescope in its
Warm Spitzer mode.
From these measurements and
catalog optical photometry we derive albedos and diameters
of the observed targets.
The overall goal of our ExploreNEOs program is to study
the history of near-Earth space by deriving the physical
properties of a large number of NEOs.
In this paper we describe both the scientific
and technical construction of our ExploreNEOs
program. We present our observational, photometric,
and thermal modeling techniques. We present results
from the first 101~targets observed in this program.
We find that the distribution of albedos
in this first sample is quite broad, probably 
indicating a wide range of compositions within the
NEO population.
Many objects smaller than one kilometer
have high albedos ($\gtrsim 0.35$), but
few objects larger than one kilometer
have high albedos. This result is consistent
with the idea that these larger objects are
collisionally older, and therefore possess surfaces
that are more space weathered and therefore darker,
or are not subject to other surface
rejuvenating events as frequently as smaller
NEOs.
\end{abstract}



\keywords{minor planets, asteroids --- infrared: solar system ---
surveys}


\section{Introduction}

\subsection{Near Earth Objects}

The majority of Near Earth Objects (NEOs) originated in collisions between bodies
in the main asteroid belt and have found their way into near-Earth space via
complex dynamical interactions.
This transport of material from the
main belt into the inner Solar System has shaped the histories of the terrestrial planets.
Together with comets,
NEOs have delivered materials such as water and organics essential for the development of life, and they offer
insight into both the past and future of life on Earth. 
The close match between the distribution
of small NEOs and the lunar crater record
\citep{strom05}
demonstrates that impacts of objects from near-Earth
space are common.

The NEAR-Shoemaker and Hayabusa space missions to NEOs
returned a wealth of information on two very different examples
of our near-Earth neighbors, and other NEO space missions
are under study.
However,
despite their scientific importance, key characteristics of the
NEO population --- such as the size distribution, mix of albedos and mineralogies, and contributions from so-called dead or dormant comets ---
remain largely unexplored, especially in the size range below 1 km
\citep{stuartbinzel}.
Critically, recent evidence (\S2.1.1) suggests that the size distribution
of NEOs may undergo a transition at $\sim$1~km,
and that the smaller bodies
may record fundamental
physical processes that are presently occurring in the 
Solar System but not understood.


While the rate of discovery of NEOs has risen dramatically in recent years, efforts to understand
the physical characteristics of these objects lag far behind. At
present there are almost 7000~NEOs known.
The WISE mission will discover hundreds of NEOs
\citep{mainzer2006,mainzer2010}, and
the Pan-STARRS program is likely to increase the number of known NEOs
to $\sim$10,000
\citep{kaiser2004}.
However, the number of NEOs with measured physical properties (albedo,
diameter, rough composition) is less than~100
\citep[e.g.,][and references therein]{wolters}.
Almost 99\% of all NEOs remain
essentially uncharacterized.

\subsection{Warm Spitzer}

The Spitzer Space Telescope \citep{werner04} was launched in August 2003,
and carried out more than five-and-a-half years  of
cryogenic observations at 3.6--160~microns. On 15~May~2009,
the onboard liquid helium cryogen ran out, making additional long wavelength
observations impossible. However, because of the thermal design of the 
the cryogenic telescope assembly and the low thermal radiation environment of 
its solar orbit \citep{gehrz07}, the telescope and instruments
are passively cooled to approximately 
26~K and 28~K, respectively. This allows for continued
observations at the two shortest wavelengths of
3.6~microns (CH1) and 4.5~microns (CH2) with the IRAC camera \citep{fazio04}.
In Spring, 2008, NASA approved a two year extended ``Warm
Spitzer'' mission whose focus would primarily be on a small number of 
``Exploration Science'' projects of 500~hours or more.
We showed in \citet[hereafter T08]{trill08} that Warm Spitzer
could successfully be used to derive the albedos and 
diameters of NEOs. In late 2008
we were awarded 500~hours of Warm Spitzer time to study
NEOs in the ExploreNEOs program, described here.

\subsection{Results from the first 101 targets}

Our observations will be carried out between
July, 2009, and July, 2011.
In this paper we present a description of our
overall program, including target selection,
observation planning, and scheduling 
constraints (\S2). We present observations
and measured fluxes for the first 101~targets
observed in our program (\S3). In \S4, we 
describe the thermal models that we use to generate
derived albedos and diameters from measured
fluxes. We then describe
results from the first 101~targets in our program
(\S5).
Finally, we describe future directions for this project
in \S6.
This paper is intended to serve as a global
reference for all subsequent papers produced
from the ExploreNEOs project. In particular,
Section~2 (program definition), Section~3 (observations
and flux measurements), and Section~4
(thermal modeling) apply not just to these first
101~targets, but to the entire project and to
future papers.

    %


\section{ExploreNEOs: The Warm Spitzer NEO Survey}

\subsection{Science goals}

The primary science goal of the ExploreNEOs program is to explore
the history of near-Earth space. We do this by studying both the
global
characteristics of the NEO population as well as the physical properties
of individual objects.

\subsubsection{The size distribution of small NEOs}

The size distribution of a population of small bodies
records the evolution
of that population.
Most NEOs are remnants of main belt asteroids
that have reached their current location via the following
sequence of events: (i) Asteroids collide in the main belt and
create fragments; (ii) the fragments drift in semimajor axis
across the main belt over hundreds of millions of years by the
sunlight-driven non-gravitational thermal force called the
Yarkovsky effect; and (iii) they eventually reach chaotic
resonances produced by planetary perturbations that can push them
out of the main belt and into the terrestrial planet region
\citep{bottke06}. The
NEO population is also comprised of numerous dormant and active
Jupiter-family comets, many of which originated in the Transneptunian
region \citep{bottke02}.  In much the same way that rocks in a streambed
suggest the nature of events upstream, the NEO population
provides us with critical clues that can tell us about the nature
and evolution of the source populations (here, the main belt and
Kuiper belt).

Recent work suggests that
the orbital and size distributions of NEOs may change dramatically
from km-sized bodies to sub-km bodies.  The evidence comes
from a range of sources: (1) New work shows that the thermal spin-up mechanism called YORP
causes many small NEOs to shed large amounts of mass over
timescales much shorter than their dynamical lifetime
\citep[e.g.,][]{bottke06,walsh08},
thus modifying the size distribution
of NEOs.
(2) A paucity of observed small comets as well as lack of small
craters on young surfaces like Europa indicates that the
Jupiter-family comet population may be highly depleted in objects
compared to our expectations for collisionally-evolved
populations \citep{bierhaus}.
If true, sub-km comets essentially do not exist or
some highly efficient (and unknown) mechanism eliminates them
prior to reaching the vicinity of Jupiter (and possibly Saturn).
Probes of the size and albedo distributions of sub-km NEOs
will provide critical constraints.
(3) Young asteroid families residing near important main belt
resonances (e.g., 3:1~mean-motion resonance with Jupiter) may supply many more
sub-km asteroids than km-sized asteroids
\citep{vernazza08,nesvorny09}.
Precise measurements
of the 
size distribution of sub-km NEOs will place strong constraints on
these processes.
(4) Preliminary debiased observations from Spacewatch indicates
many sub-km asteroids simply do not survive long enough to reach
$a<2$~AU orbits (J. Larsen, pers.\ comm.).

\vspace{1ex}

These lines of evidence suggest that the behavior and
evolution of sub-km NEOs may be very different from that
of km-sized NEOs.
Precise measurements of the size 
distribution of NEOs will place important constraints on
the interplay among the processes described above;
at present, there is no large-scale
empirical dataset to serve as a benchmark against which
theoretical models can be tested.

\subsubsection{The search for dead comets}

A long-running debate concerns the fraction of the NEO population
that has a cometary origin: How many NEOs are ex-comets that
have exhausted all of their volatiles and now appear indistinguishable
from asteroidal bodies? 
This question has important consequences for the history
of near-Earth space as well as for the history of life 
on Earth, as comets carry significant volatile and organic
material.
Since the dynamical lifetimes of comets
generally exceed their active lifetimes, there are expected to be a
large number of dormant or extinct comets that are catalogued as asteroids.
\citet{fernandez05}
estimate that
some 4\% of all NEOs are ``dead'' comets, while
\citet{binzel04}
estimate that up to 20\% of the NEO
population may be extinct comets, after correcting for observational
bias against the detection of low (``comet-like'') albedo objects in the known population of NEOs.
However, caution should be exercised in considering these results as they are
based on small number statistics and largely exclude the size range below 1~km.

\begin{figure}
\includegraphics[angle=270,scale=0.5]{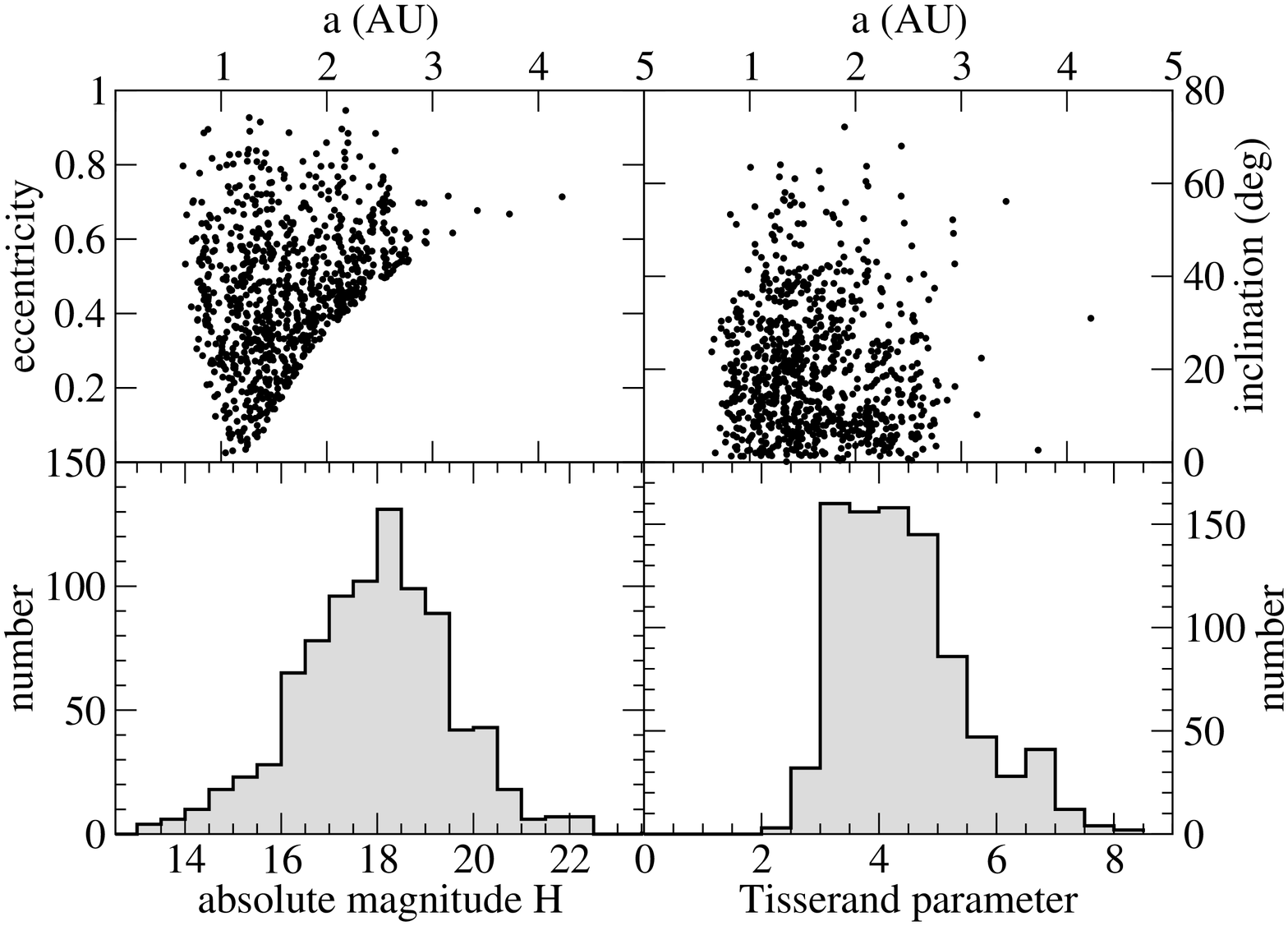}
\caption{Our pool of target objects,
showing orbital elements,
Solar System absolute magnitude $H$, and
Tisserand parameter with respect to Jupiter.
Some 5\% of our targets
have $T_J<3$, suggestive of bodies
that may have originated in the outer
Solar System.
\label{fourpanels}}
\end{figure}

In the ExploreNEOs program
we will derive albedos for a large number of 
small NEOs.
In doing so, we will measure the fraction
that may be dead comets.
Unlike optical surveys, we are quite
sensitive to low albedo objects due to their larger
ratios of thermally emitted to solar reflected radiation
in both IRAC CH1 and CH2.
Of course, most dark asteroids may not be dead comets; we
will be guided by orbital parameters and dynamical considerations in identifying objects that
may be of cometary origin (Figure~\ref{fourpanels}).
Since dead comets are dynamically linked to the outer
Solar System, just as most NEOs are links to the evolution
of the main asteroid belt, we will be using NEOs as a probe of the 
evolution of the small body populations of the Solar System.

\subsubsection{NEO origins and compositions}

The majority of NEOs are believed to be S class asteroids, a rocky and relatively
volatile poor asteroid type generally found
in the inner main belt \citep{stuartbinzel}.
The majority of main belt asteroids are C type asteroids,
more volatile and organic rich and generally
found in the outer main belt \citep{gradie,carvano10}.
In general, 
S asteroids have moderately high albedos ($\gtrsim$0.15), and C asteroids have low
albedos ($\lesssim$0.10). The interplay between the source regions, 
dynamical paths, and 
overall evolution of main
belt asteroids to NEOs is complicated
\citep[e.g.,][]{wisdom82,froeschle,bottke02}.
Measuring the albedos of a large
number of NEOs
will allow us to determine
the
relative mixing fractions of
main belt asteroids and outer 
Solar System objects
in near-Earth space.



We will also measure the albedo distribution
as a function of size.
As described above, the relative fractions of 
inner and outer Solar System objects may differ
for small (sub-km) and larger (km-sized) NEOs.
Precise measurements of the albedo distribution
as a function of size will constrain the evolution
of those very small bodies.

\citet{delbo03} found that the albedos for S-class (and related
classes) NEOs
rise from their main-belt average value of around~0.22 to greater
than~0.3 for objects smaller
than 500~m. \citet{harris06} re-examined the trend and found
that while detection bias against small dark objects may contribute,
it is unlikely to be the sole explanation.
Results from our Spitzer pilot
study appear to confirm
this trend (T08),
though with small numbers and
not insignificant error bars.
Smaller objects should have statistically younger surfaces
(more likely to have suffered a recent surface-refreshing event), so
change in albedo with size could be an indication of space weathering
processes.  These processes have been
studied for S-types
\citep[e.g.,][]{hapke01,chapman04}
but are much
less well understood for the darker C-types,
for which we will
derive excellent albedos (due to their darker surfaces,
as described above).

\subsubsection{The physical properties of individual NEOs}

Most airless bodies in the Solar System are covered to some degree with regolith,
layers of pulverized rock that are produced over time by collisions with both large and
small bodies.
Our Warm Spitzer NEO survey will take us into a size regime in which
very little is known about asteroid regolith properties. It has
been suggested, based on indirect evidence, that bodies
smaller than 5 km may be nearly devoid of fine regolith
\citep{binzel04,cheng}, but recent thermal
observations apparently do not confirm this
expectation \citep{harris05,harris07,delbo07,muellerthesis}.
The small (500 m long) NEO Itokawa was visited by the Japanese spacecraft Hayabusa, and was found to have a highly varied surface, with both regolith-free regions and regions of substantial regolith. This unexpected result has greatly increased interest in the surface properties of NEOs and how they depend on an object's history. Furthermore, the presence or absence of regolith strongly influences surface thermal inertia, 
which is a measure of the resistance of a material to temperature changes (e.g., diurnal cycles) and which is
a key parameter in model calculations of the Yarkovsky effect, which causes gradual drifting of NEO orbits and is therefore an important dynamical effect. 
%

While our Warm Spitzer observations will not allow us 
to measure thermal inertia directly,
indirect information can be gained from the distribution in apparent
color temperature, which is determined from the thermal flux ratio at
IRAC wavelengths.  The thermal flux ratio will be measured
with high significance
in the case of low-albedo targets, for which contamination
from reflected sunlight is much reduced.
We will derive the 
average thermal inertia of our target sample ---
and therefore information on the absence or presence
of regolith ---
from a statistical analysis of
the color-temperature distribution.
\citet{delbo07} have used this
method, using a much smaller data base of ground-based observations, to
determine the typical thermal inertia of $\sim$1~km NEOs.
Our ExploreNEOs program will allow us to determine for the
first time the typical thermal inertia of sub-km NEOs.
This result will be crucial for modeling the Yarkovsky
effect, as the strength of the Yarkovsky effect depends
sensitively on thermal inertia \citep[e.g.,][]{bottke06}.
Additionally, understanding --- and potentially mitigating --- the
Earth impact hazard requires detailed analysis of the thermal inertia and consequent
Yarkovsky effect on NEOs
\citep{giorgini2002,milani2009}

Perhaps 15\% of NEOs are binaries \citep{pravec2005}.
While Warm Spitzer will not be 
able to resolve binary NEOs, thermal
measurements of NEOs that are known to be binary
or that turn out to be binaries will yield densities (assuming
that the members of the binary share a common albedo and density), which in turn
suggest compositions and internal strengths. 
(Reasonably high quality orbital periods and semi-major
axes are necessary to derive densities.)
This is the only way to measure the internal properties of
non-eclipsing asteroids short of visiting them with spacecraft.
In this program,
we are likely to measure fluxes for tens of NEOs
that will turn out to be binaries,
enabling future work that constrains the evolution of those
bodies and, by proxy, the entire NEO population.

\subsubsection{Impact of our results: The Congressional mandate and public awareness}

The U. S. Congress has mandated that 90\% of all NEOs larger
than 140 meters in diameter be identified by 2020 using a combination
of ground-based and space-based facilities.  As
discussed by the Science Definition Team
report
to Congress\footnote{{\tt http://neo.jpl.nasa.gov/neo/report2007.html}},
this retires 90\% of the hazard
posed to the Earth by asteroid impacts.
Thermal observations of the NEO population will
allow us to derive the true NEO size distribution,
which the optical surveys do not: they measure the
brightness distribution, 
which cannot be converted to a size distribution
easily because the albedos of the NEOs are 
unknown.
Knowledge of the size distribution is critical for
estimates of the Earth impact hazard.
Increased understanding of radiation forces
is important both because 
small bodies undergo orbital evolution due to the
Yarkovsky effect,
but also because radiation forces have been proposed
as a mitigation technique. In both cases, a more
complete understanding, based on observations,
is necessary.
%

\subsection{Design of the program}

\begin{figure}
\includegraphics[angle=270,scale=0.5]{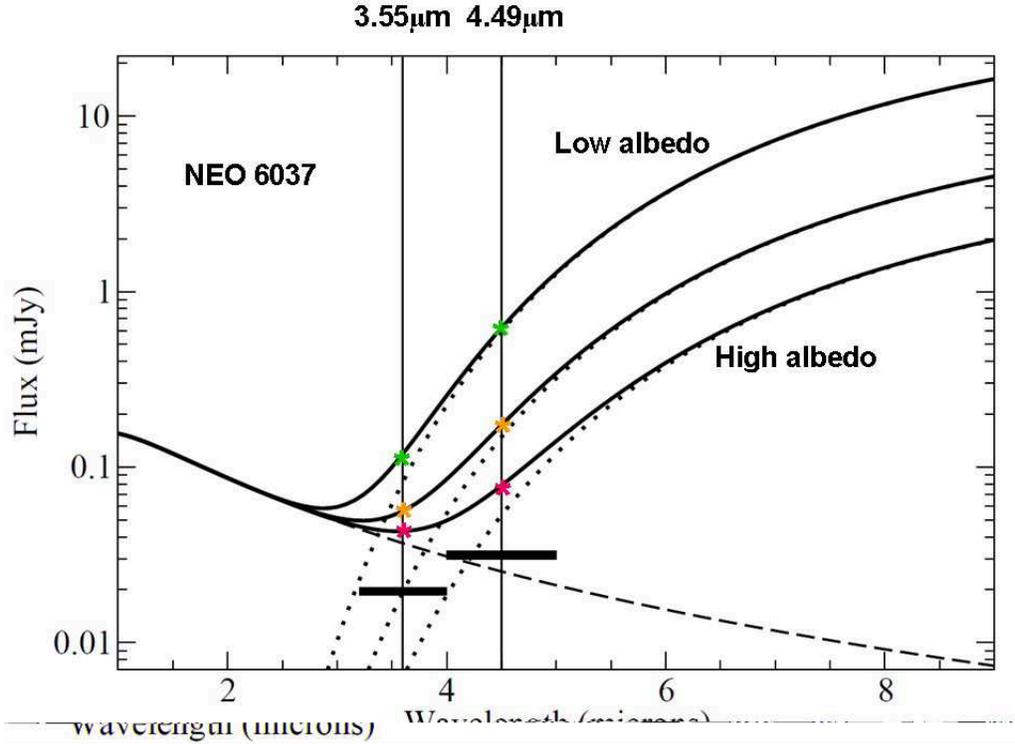}
\caption{Spectral energy distribution (SED) of an example NEO
observed in T08
(middle solid curve, for an albedo of~0.37). 
Also shown are hypothetical SEDs for
this object if it were to have a low albedo
(\pv\ of~0.1, top curve) or high albedo (\pv\ of~0.6, bottom curve)
with the $H$ value held constant. The dashed curve shows the
reflected light component and the dotted curves show the
thermal components. The combined fluxes are given by the solid curves.
The thick horizontal lines indicate the approximate sensitivity
levels at the Warm Spitzer wavelengths applicable to our program. The
stars indicate the points on the combined flux curves at
which the fluxes are measured; the vertical distances between these points
and the dotted lines indicate the extent to which reflected sunlight
contributes to the flux measurements. The problem increases with
increasing albedo. The method for correcting for reflected sunlight
is described in Section 3. \label{sed}}
\end{figure}

NEOs have relatively hot dayside surface temperatures ($>$250~K). Fluxes in IRAC CH1 and CH2
will therefore contain
large thermal flux components (Figure~\ref{sed}). Sizes of NEOs
can be derived 
from their thermal measurements. These 
size determinations can be combined with reflected light
data 
(visible magnitudes obtained from ground-based observations) 
to derive albedos.

\subsubsection{Sample selection and subsamples}

Our target selection process is as follows.
We began with the list of all known NEOs as of July, 2008 ($\sim$5000~bodies).
For each of these, we calculated the dates 
when the target is within the Spitzer
visibility zone (solar elongations 82.5--120~degrees).
We further culled to retain only those objects
with small positional uncertainties ($<$150~arcsec)
as seen
by Spitzer during those times
to ensure that these targets will fall within
the IRAC field of view.
We applied a cut to ensure that no object
is moving faster than the Spitzer tracking
rate of 1~arcsec/sec (no presently
known NEOs are excluded by this cut).
The remaining list is our pool of targets.
This pool includes NEOs that span a range
of sizes, orbits, and (presumably)
physical properties (Figure~\ref{fourpanels}).

\begin{figure}
\includegraphics[angle=270,scale=.50]{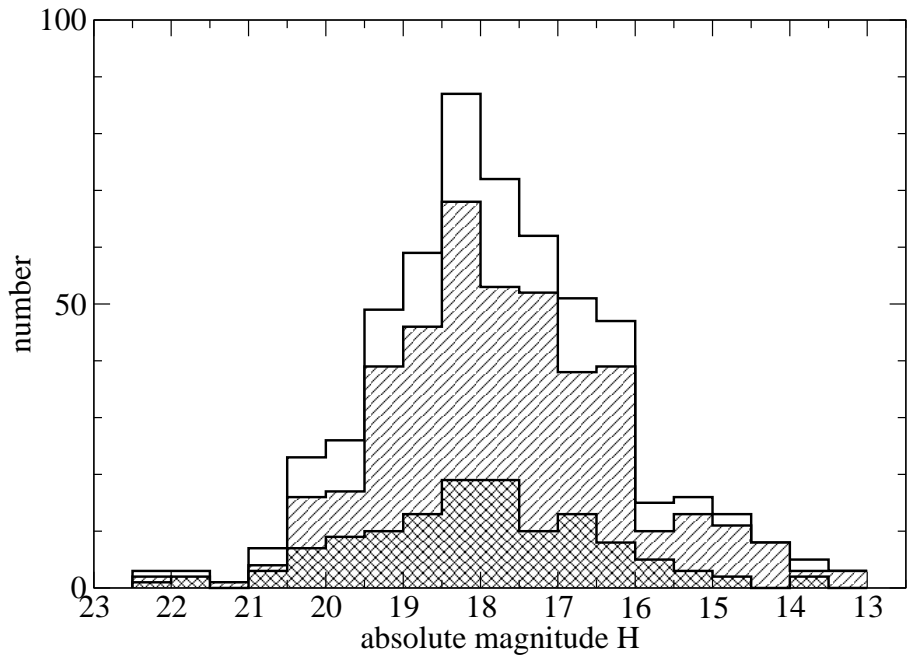}
\caption{The distribution of $H$ (Solar System
absolute magnitude) for our targets.
Shown here are histograms for 
the entire sample
(tall clear histogram);
all targets
with $t_{int}<2000$~sec
(upper right to lower left hatching);
and
all targets with
$t_{int}=2000$~sec
(upper left to lower right hatching).
All histograms count from the bottom,
that is, there are $\sim$50~targets
in the ``not-longest'' (upper right
to lower left) and 10~targets in
the``longest''
(upper left to lower right)
bins at $H$=17.5--18.
The distributions for ``not-longest''
and ``longest'' are very similar, which means
that observing fewer objects at the longest
integration times does not introduce
a significant bias.
\label{Hhist}}
\end{figure}

We divided these targets into two subsamples in order to 
maximize scientific returns and limit required telescope
time. The first pool, which we call the ``certain''
pool (PID~60012, where the PID is the unique
Spitzer Program identification number; 584~targets), contains all targets that do not have the longest
exposures times (see below), as well as all targets with
Tisserand parameters with respect to Jupiter less than~3.1.
(These low Tisserand parameter values are thought
to suggest a higher likelihood of an object's origin
in the outer Solar System \citep{levison}, as opposed to the main asteroid
belt. We use a cutoff of~3.1 instead of the nominal
cutoff of~3.0 in 
order to be able to probe any effect completely.)
The second pool (PID~61013; 77~targets) contains all remaining targets with 
the longest integration times. Around 70\% of these targets
will be observed as a statistical sample.
We note that 
because of the large variation in flux
for a given NEO over its visibility window,
choosing the brightest targets at any
given time does not bias against
targets with the largest $H$ values (Figure~\ref{Hhist}).

We have three more subsamples in our ExploreNEOs program.
The ``ground-truth'' targets (PID~61012; 2~targets) are NEOs
that were not already included in our sample and about which
we know albedo and diameter through some independent set of
measurements (spacecraft, radar, etc.), and which will be used
as calibrators for our larger program. We also have a sample
of multi-visit targets (PID~61011; $\sim$10~targets), 
which will be observed over
a range of observational geometries in order to assess systematics
in our thermal modeling. Finally, we have a pool of generic targets
(PID~61010; 25~targets), which are observations that are planned but whose
targets are unspecified at this time. These observations will be used
to observe NEOs are that have been discovered (or whose orbits
have been refined) since our primary
target list was fixed.

\subsubsection{Observational strategy}

Since the fluxes for NEOs change dramatically
on a daily basis, the most efficient way to
carry out this program would be to have 
a different planned observation for each day of the Warm 
Spitzer mission, 
for each target, with integration times 
carefully customized for the predicted
brightnesses. However, neither our
team nor the Spitzer Science Center could realistically deal
with this explosion of candidate observations.
Instead, we adopted a set of template observations
with integration times of [100, 400, 1000, 2000]~seconds per
channel.
We reject observations which would require integration times of more
than 2000~sec
to reach the required SNR.
All generic targets are assigned the 2000~sec
integration time, our longest.

We set our minimum detection threshold at SNR$\geq$15
as of 2010~Jun~1
(replacing
the SNR$\geq$10 requirement used prior to that date).
We calculate the smallest exposure time for
which a given target is visible to Spitzer
with our minimum detection threshold for
at least five consecutive days; this five day
window aids in scheduling our observations.
Thus, a target that is bright enough to be
observed with a short integration time for only a few 
days is assigned a longer integration time in
exchange for a more generous timing constraint.
This improves the overall schedulability
of our program substantially while
still grossly tailoring AORs to predicted target fluxes.
(A Spitzer AOR is an Astronomical Observing
Request --- essentially, a single planned observation.)
The final result is that each target is assigned
a single integration time for the duration of
the Warm Spitzer mission.
There are 
[309, 149, 101, 25]~targets that
have integration times
of [100, 400, 1000, 2000]~seconds in PID~60012.
(All AORs in the other four PIDs have 2000~seconds
integration times.)


We use
the moving cluster
AOT (Astronomical Observation Template --- a
fixed observing pattern used by Spitzer),
tracking according to the standard NAIF ephemeris.
We eliminate all potential observations near
the galactic plane because of background source confusion.
Our dithered observations alternate
between the bandpasses during the observation
to reduce the
relative effects of any lightcurve variations within the observing period, and
to maximize the relative motion of the asteroid to help reject
background sources; this technique has been 
validated in T08.
For observations that have sufficient
SNR in the individual exposures (or some binning of frames), we check
for variations
over the duration of the AOR.

\subsubsection{Predicted fluxes}

%
Our flux predictions are based on the Solar System
absolute optical magnitude $H$, a measure of $D^2$\pv, as reported by \emph{Horizons}.
(Here $D$ is the diameter and \pv\ is the 
geometric albedo, or reflectivity).
$H$ values for NEOs are of notoriously low quality and tend to be skewed towards too bright values 
\citep{juric02,rt05,parker08}.
We therefore assume an $H$ offset ($\Delta H$) of [0.6, 0.3, 0.0]~mag
for [faint, nominal, bright]~fluxes, respectively. That is, we
hypothesize
that the nominal true magnitude of a given object is 0.3~mag fainter than
the \emph{Horizons} value, within a range 0.6--0.0~mag
fainter than the \emph{Horizons} value.
Reflected light fluxes are calculated from $H+\Delta H$ together with the observing geometry and the solar flux at IRAC wavelengths. 
Nominally,
asteroids are assumed to be 1.4 times more reflective at IRAC wavelengths than in the V band
\citep[T08,][]{harris09}.
Thermal fluxes also depend on \pv\ ($D$ is determined from $H$ and
\pv) and  $\eta$,
a model parameter that attempts to capture details of the physical
properties (rotation rate, surface roughness, etc.) of the asteroid.
We assumed \pv= [0.4, 0.2, 0.05]~for the [faintest, nominal,
brightest]~thermal fluxes, that is, we hypothesized that
asteroid albedos are in the range 0.05--0.4 \citep{tedesco_simps,binzel_asteroids3}.
The nominal $\eta$ value is determined from the solar phase angle
$\alpha$ using the linear relation given by \citet{wolters};
0.3 is [added, subtracted] for [faint, bright] fluxes to capture the scatter in the empirical relationship derived
in Wolters et al.

From this range of parameters, we calculate the range
of predicted fluxes in the two IRAC bands for each day
of the two year Warm Mission.
The resulting thermal fluxes are convolved with the posted IRAC
passbands\footnote{\url{http://ssc.spitzer.caltech.edu/irac/calibrationfiles/spectralresponse/}} to
yield predicted fluxes.
%
%
%
We then calculate the integration time required
to reach the required SNR for the faintest predicted
flux for each target for each day.

\subsubsection{Scheduling}

Each of our targets, with few exceptions, has at
least
one acceptable visibility window of at least five days.
A few targets have no available visibility windows of
five days or longer; for these we take the longest available
visibility window. Since the nature of our program is 
that of a large sample in which individual measurements
are less important than the entire suite of results,
if a small number of observations fail, either through
scheduling issues or data quality issues, the overall
impact of this program is not affected.
With $\sim$700~targets to
observe during the two year Warm Mission, nominally we expect
to have one target observed per day, on average, and our yield to
date has been slightly higher than this expected rate.

\section{Observations and flux measurements}

We report here observations made during
the period 28~July~2009 through 4~November~2009
(that is, from the
IRAC warm instrument characterization, IWIC,
campaign through the end of 
the seventh post-cryo [warm] campaign, PC007).
%
The data were reduced using the IRACproc software \citep{schuster06}, which is
based on the {\tt mopex} routines provided by the Spitzer Science Center \citep[SSC;][]{makovoz05}.  The 
Basic Calibrated Data (BCD) produced by version 18.12.0 of the pipeline were used in the reduction.
Mosaics of each AOR were constructed using the ``moving object mode,'' which aligns the individual
images in the rest frame of the moving target, based on its projected motion.  The outlier rejection
in the mosaicking process then removes or minimizes the fixed background objects in the field and any transients due to
cosmic rays or array artifacts.
The images are rebinned to a pixel scale of
0.8627~arcsec/pixel in the final mosaics.
We extracted the photometry using the {\em phot} 
task in IRAF.  The noise in the image was estimated in the region near the NEO, and the extraction
used an aperture radius of 6~mosaic pixels (5.1762~arcsec), with a sky annulus with inner radius of 6~pixels and 
outer radius of 12~pixels.  These parameters are significantly smaller than the aperture size of 10~instrumental
pixels (12.2~arcsec) used in the IRAC calibration measurements \citep{reach05}.  The 
smaller aperture was chosen to reduce the effects of background objects in the aperture in 
crowded fields where some of the NEOs were observed.  In order to calibrate our NEO photometry, we 
extracted the photometry of IRAC calibration stars observed in the same campaigns as our NEO observations, at the same
detector temperature and bias settings. Our
zero points in channels 1 and 2 were then adjusted to make the standard star magnitudes match those reported
by \citet{reach05}.  For the observations taken after the final IRAC warm mission biases and temperatures
were set on 24~September~2009, the zero points used are 17.864~and
17.449~magnitudes for the 3.6~and 4.5~micron bands, 
respectively.
The measured fluxes and associated errors are reported in
Table~\ref{datatable}.

The errors in the fluxes are from the output of the {\em phot} routine,
which include the instrument parameters such as gain and read noise, and
the noise in the images, which includes residuals from the
incompletely-rejected background sources.  In addition to these errors, the
Spitzer Science Center
reports\footnote{http://ssc.spitzer.caltech.edu/irac/documents/iracwarmdatamemo.txt}
that the calibration of the Warm Mission data is preliminary, and the
pipeline reduction could have errors of up to 5\%-7\% at 3.6 $\mu$m and
$\sim$4\% at 4.5 $\mu$m, primarily from errors in the linearity correction.

\section{Thermal models and discussion of errors}

\subsection{Color corrections and the reflected light component}

Due to the width of the IRAC passbands, our measured flux values must be color corrected.  Also, the observed asteroid flux contains reflected sunlight which must be subtracted before thermal models can be applied. 
Due to their different spectral shapes, different color corrections apply to the thermally emitted and reflected flux components; color corrections for the latter are negligible.
Our approach is to estimate the reflected flux component for both
IRAC wavelengths. We then subtract these reflected
light components from the measured fluxes to derive the uncorrected
thermal fluxes. Finally, as described below, we color-correct these
thermal fluxes assuming a thermal asteroid spectrum.

The flux component from reflected sunlight was assumed to have the spectral shape of a 5800~K black body over IRAC's spectral range.  The flux level was determined from the solar flux at 3.6 micron \citep[$5.54\times10^{16}$~mJy;][]{solarspec}, the solar magnitude of $V=-26.74$, and the asteroid's $V$ magnitude as determined from the observing geometry and the known $H$ value.
Reflected fluxes were multiplied by 1.4 to account for the increased reflectivity at 3.6~micron relative to the V band \citep[e.g., T08,][]{harris09}.

Color-correction factors for the thermal flux were determined using the method described in \citet{mueller07} and T08, that is, by convolving the thermal
spectrum with the IRAC bandpasses measured in flight.
For the thermal spectrum, we assumed that \pv\ is a function
of phase angle (see discussion below) and that $p_V=0.1$; varying $p_V$ within reason leads to fluxes changes of less than 1\%.
Our derived color-correction factors scatter around~1.17 and~1.09 for CH1
and CH2, respectively, where physical flux is equal to
in-band fluxes divided by the color-correction factors.

\subsection{Thermal models}

In almost all cases, the 4.5~micron flux is dominated by
thermal emission; in many cases, the 3.6~micron flux is
not (Figure~\ref{sed}). Therefore, our 
primary analysis is
driven by the 4.5~micron flux. The 3.6~micron fluxes
are included in the analysis, but at low weight;
these CH1 fluxes generally make no distinguishable contribution to
the solutions.

We derive the diameter and geometric albedo of NEOs by combining thermal measurements with optical photometry (from ground-based observations), using a thermal model.  A suitable model for NEOs is the Near-Earth Asteroid Thermal Model
\citep[NEATM; ][]{neatm},
where
thermal fluxes are determined by integrating the Planck function over the illuminated and visible portions of a sphere.
The NEATM incorporates a variable adjustment to the model surface temperature through the parameter
$\eta$, allowing a 
correction for the thermal effects of shape, spin state, thermal inertia, and surface roughness, and enabling the model and observed thermal continua to be accurately matched. 
More detailed thermophysical modeling \citep[e.g.,][]{harrislagerros}
would require knowledge of shape and spin state, which is generally unavailable
for our poorly studied targets.

Throughout this paper, we assume a slope parameter $G$ (in the HG
system) of~0.15, as is customary for asteroids,
unless otherwise stated;
this value of~0.15 is used (for example) by
the Minor Planet Center. $G$ is needed for determining the
expected V magnitude at the time of our Spitzer observations, a
prerequisite of our correction for reflection sunlight.
Also, $G$ sets the value of the phase integral (0.393), the ratio
between bolometric Bond albedo (needed for determining the temperature)
and \pv.
We expect that uncertainty in $G$ is only a small source
of uncertainty in the final determination of albedo and diameter
for our Warm Spitzer data; this will be explored in future
work (Harris et al., in prep.).

For about 20\% of our targets NEATM fits to the IRAC CH1 and CH2
fluxes and optical
magnitude (derived from the \emph{Horizons} absolute magnitude $H$)
provide reasonable results for the three unknowns (diameter, albedo,
and $\eta$).
(For our nominal results presented in Table~\ref{datatable}
we use the \emph{Horizons} $H$ values even if
other values are known in the literature, for 
consistency. Future papers
will explore the implications of this and substitute improved
$H$ magnitudes derived from other sources, including
new measurements made by our team.)
With only three data points and three unknowns we cannot use standard
goodness-of-fit techniques to estimate uncertainties;
in any case we expect our overall modeling uncertainties to be far
larger than the formal uncertainties derived from the error bars on
the
flux measurements. An analysis of overall uncertainties, including a
comparison of our results with published sizes and albedos where
available,
will be the subject of future work (Harris et al., Mueller et al., in prep.).
For our present purposes we simply filter our two-channel
NEATM fits to accept only those cases
in which the
CH2/CH1 thermal flux ratio is consistent with our expectations
for NEOs (see \S5.1).
The mutual consistency of these two modeling approaches is discussed
in
Section 5.1.

In most cases we are unable to obtain reasonable solutions fitting
NEATM to the CH1 and CH2 fluxes
due to the large and uncertain contribution of reflected solar
radiation in CH1.
Our compromise is to use an empirical relationship between $\eta$ and
solar phase angle
\citep{delbo03,delbo07,wolters}. In T08 we demonstrated that this
technique gives results
that are in reasonable agreement with a NEATM fit to all four
cryogenic
IRAC bands.

\subsection{Errors and uncertainties}

Some NEOs have rotational flux variability in excess of 1~magnitude.
Therefore, in addition to the systematic $H$ uncertainties
described above, additional uncertainties can be introduced
from target lightcurves. For faint targets, lightcurve-induced
uncertainties are minimized
(see Appendix~A of T08), but for bright targets, our model
results may include lightcurve-induced errors;
an example of this, Eros, is described below.

We have assumed that all asteroids are 40\% more reflective at
3.6~and 4.5~microns than at V~band.
Additional ground- and space-based data (as our
sample grows) will allow us to refine this assumption.
However, varying this reflectance ratio between~1.0
and~1.7 produces changes in diameter of $<$5\%
(corresponding to albedo changes $<$10\%) in almost
all cases, with the exceptions (few percent) being cases with 
particularly low SNR data.

\citet{wright07} has tested the NEATM against a sophisticated thermophysical model and finds that it gives diameter estimates that are accurate to 10\% for phase angles less than 60~degrees,
even for the non-spherical shapes typical of NEOs. Including all sources of errors discussed in T08, we estimate that the total uncertainties in
our modeling will be $\sim$20\% in $D$ and $\sim$40\% in $p_V$.
Uncertainty in $H$ ($\sim$0.3~mag;
\citet{juric02,rt05,parker08})
adds 30\% to the error budget in $p_V$,
leading to a total albedo uncertainty of 50\%, but leaves $D$ practically invariant
\citep{harrisharris}.
We emphasize that future ground-based work,
such as Pan-STARRS and our supporting
ground-based campaign,
will provide much improved $H$ values.
Finally, we stress that the accuracy of our diameter
and albedo results
for the scientifically valuable sub-set of low-albedo NEOs will be significantly higher than the overall estimates given above, due to
high SNR thermal flux measurements in both
IRAC bands.

The previous discussion addresses systematic uncertainties in our
model
results. We also can estimate our repeatability, which traces
random uncertainties that may arise in flux measurements,
for example. There are three targets to date that have been
observed multiple times: each of 2003~WD158, 2004~JX20,
and 2006~LF were observed both during IWIC and during
nominal post-cryo missions. For each of these six independent
sets of measurements we derive albedo and diameter as
described above, and we can compare the results for each
pair of observations to estimate repeatability uncertainties.

Both solutions for each of these three targets are presented in 
Table~\ref{datatable}. In all three cases, the two diameter solutions agree
to within 15\%.
Each observation in a pair was made at different phase angles
(and hence we use different $\eta$ for each solution in the pair),
further showing that our repeatability is quite good. This
suggests that, at present, the
errors
derived from repeatability are smaller than the systematic errors
described above.

The final aspect to our total uncertainty is that of accuracy,
that is, how close are our derived model values to actual
values measured through other techniques. We discuss
below that in some cases our derived answers
are not particularly close to previously
established values, but that in other cases
our derived results match other results quite closely. The
distribution
of errors in accuracy requires our larger sample to fully
characterize.

\section{Results for the first 101 targets}

We report here observations made during
the period 28~July~2009 through 4~November~2009
(that is, from the IWIC campaign through the end
of PC007).
108~targets in our program were observed during this
period. This includes
three targets that were observed twice: once during the 
IWIC campaign
before nominal Warm Mission observations commenced,
and once during nominal observations.
Data for four targets are not usable, though
some of these data may be recoverable
in the future:
(24761)~Ahau was too faint and confused with
background sources;
(52762) 1998~MT24 was saturated;
(89355) 2001~VS78 was too close to a bright
star in the IRAC observations; and
2004~QF1 was too faint in CH1.
This leaves a total of 101~unique targets
with good data from this time period.

\begin{figure}
\includegraphics[angle=270,scale=0.5]{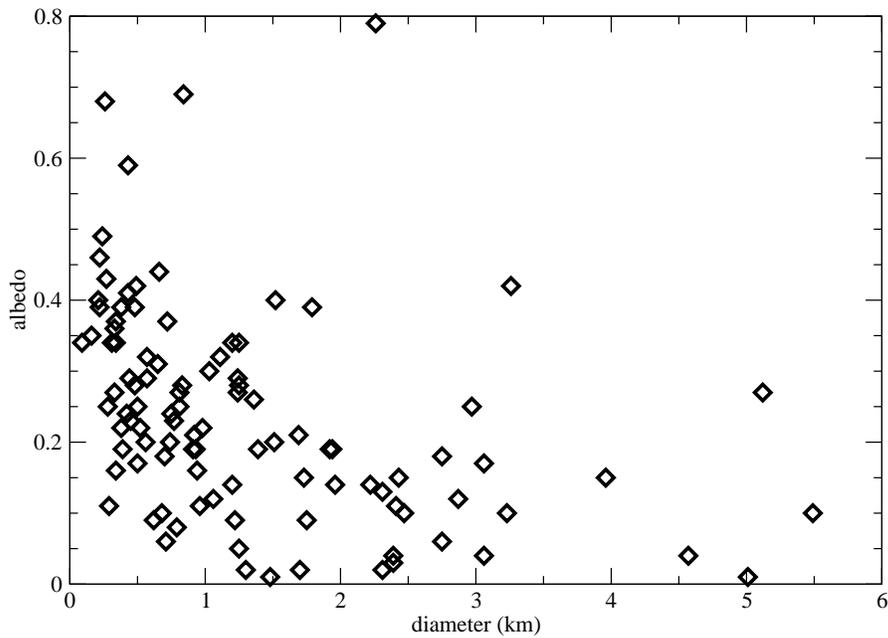}
\caption{Albedo versus diameter for our first
101~targets, using fixed-$\eta$ solutions in all 
cases. Two objects are off the plot:
42286 (2001 TN41), with modeled \pv\ of~1.03 and diameter 0.7~km,
and (433) Eros, with modeled \pv\ of~0.06 and diameter 32~km;
see \S5.3 for discussion of these objects.
The approximate errors for each data point may
be as large as 25\% in diameter and
50\% in albedo.
\label{albdiam}}
\end{figure}

We present results for these first 101~targets in Table~\ref{datatable},
which reports all the relevant information for each
of these observations: target and data-handling information;
measured fluxes and errors; observing geometries; and
derived (modeled) parameters.
The results are also presented in Figures~\ref{albdiam}
and~\ref{albhist}, and described
in the following sections.

\subsection{Floating $\eta$ and fixed $\eta$ results}

For a subset of our targets we are able to obtain apparently
reasonable results by
modeling the thermal fluxes at both 3.6 and 4.5 microns. For this
purpose we use the
full NEATM model (in combination with the optical brightness
calculated from $H$) and derive
albedo, diameter, and $\eta$, the model parameter that depends on the
thermal properties of the body.
These ``floating-$\eta$'' fits generally return more accurate results
as the properties of
each body are solved for individually. However, in the case of our
Spitzer data the ratio of
the 4.5/3.6 micron thermal flux components varies widely, probably
reflecting the significant
uncertainty associated with the correction for solar reflected
radiation at 3.6 microns
(Figure~\ref{sed}). In many cases NEATM fits fail completely or result
in
unrealistically high/low albedos and/or $\eta$ values. We find that
setting a filter in the
processing pipeline to exclude targets with 4.5/3.6 micron
thermal flux ratios outside
the range~7--10 gives reasonable floating-$\eta$ results for all cases
passing the filter.

In order to solve for albedo and diameter in the general case
(i.e., lacking a reliable 3.6 micron thermal flux value), we assume a
value for $\eta$ based on the object's observational phase angle:
$\eta = 0.013\alpha + 0.91$, where $\alpha$ is the phase angle
in degrees \citep{wolters}.
We refer to these solutions as``fixed-$\eta$'' fits. Since we do not
solve for the properties
of individual asteroids, these fixed-$\eta$ solutions are generally
less reliable (have larger instrinsic errors) than floating-$\eta$
fits to two reliable
thermal flux values.

Table~\ref{comparison} compares our derived results for the 19~targets
passing our
thermal flux ratio filter (that is, those objects
with realistic floating-$\eta$ solutions) to their respective fixed-$\eta$
solutions.
The mean fractional difference $((D_{fixed}-D_{float})/D_{float})$
between the two sets
of diameters is -7.5\%, a remarkably good
agreement, indicating that overall the fixed-$\eta$ results do not
suffer from a serious bias.
The mean absolute fractional difference
$(|(D_{fixed}-D_{float})|/D_{float})$ is 21\%.
These comparisons suggest that the additional error introduced by
using the fixed-$\eta$
approach --- the only approach possible in most cases --- is not large. For the 20\%
of our targets for which the data warrants a floating-$\eta$ solution we can gain both 
an understanding of the surface physical properties of those asteroids
and an
evaluation of the validity of our fixed-$\eta$ assumption (Mueller et al., in prep.).

\subsection{Results for the entire sample}

Figure~\ref{albdiam} shows derived albedo and diameter for our first 
101~targets.
We find that around half of our targets have diameters less than
1~km, thus increasing the number of sub-kilometer NEOs with
known properties significantly. These smallest NEOs have a wide
range of albedos, from~0.1 to~0.7, implying a wide
range of compositions.
There are no objects with
diameters less than 500~meters and albedos less than~0.1
in these initial results. This lack of smallest, darkest NEOs is an
observational bias that derives from our sample selection: we observe
only those targets that have been reasonably well observed in
ground-based (optical) programs, and such programs are biased
against small, dark objects. 

There are few objects in our
sample larger than 1~km that have albedos
larger than~$\sim$0.35 (in comparison, many subkilometer objects
have albedos this large or larger). This is consistent with the 
idea that the NEO population continues to be sculpted by collisions
(or perhaps disruptive dynamical interactions with planets), in that
the smaller objects would be younger, and thus have undergone
less space-weathering, which darkens surfaces.
The darkening timescale at 1~AU is around 1~million years
\citep{strazzulla,vernazza09}.
Thus, the dichotomy in albedos for large versus small
NEOs may indicate that many NEOs smaller than 1~km are quite young.

\begin{figure}
\includegraphics[angle=270,scale=0.5]{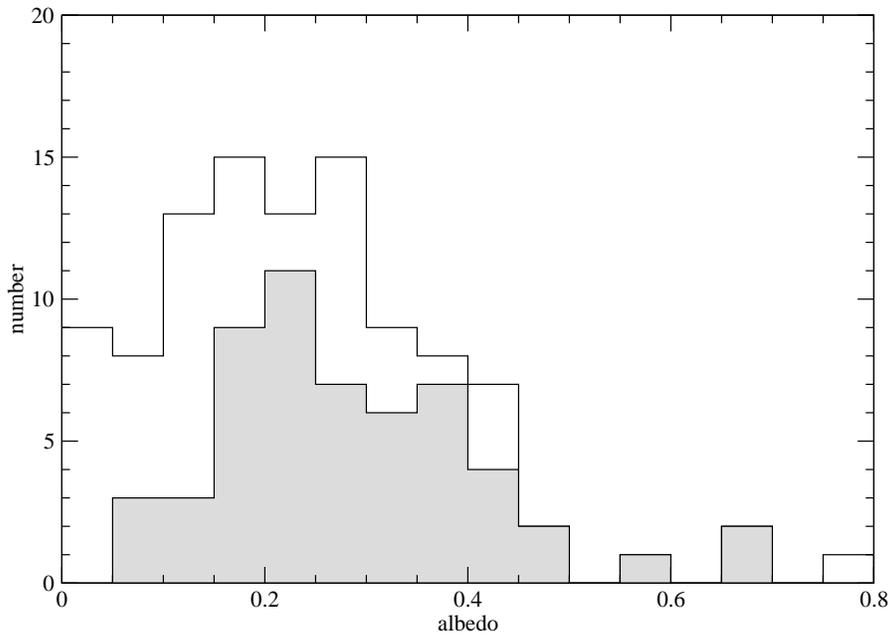}
\caption{Distribution of albedos for our first
101~targets, using model results derived
with the fixed-$\eta$ model in all cases. 
The upper line (and clear filled area) is the total for
all targets, and counts from the bottom.
The lower line (and grey filled area) is the total
for the 56~targets smaller than one kilometer. The bias
against small, dark objects is apparent.
\label{albhist}}
\end{figure}

Figure~\ref{albhist} shows the distribution of albedos for the first
101~targets. We find that the distribution is relatively
flat for albedos less than~0.3. The biases here are more 
complicated than above. Primarily, there is an optical observational
bias against dark
objects in the smallest size range.
However,
dark
objects may be more 
common in the larger size range due to space weathering.
The overall effect may be that the distribution of albedos
in the NEO population shown in Figure~\ref{albhist} may
be reasonably representative
of NEO composition diversity.
Our results here suggest that the NEO population
may be approximately equal parts bright ($p_V\gtrsim0.25$), moderate
($0.1\lesssim p_V\lesssim 0.25$),
and dark ($p_V\lesssim0.1$) objects (using somewhat
arbitrary cutoffs),
implying that a wide range of asteroid taxonomic types may be
represented.
If borne out, 
this information would trace key source regions of NEOs in the main belt.
However, given the uncertainties in our model results
described above, 
this result is quite preliminary.

Previous work \citep{morby02,stokes,stuartbinzel}
predicted that bright S-type asteroids make
up 50\%--80\% of the NEO population, with 
dark C-type NEOs being the remaining minority.
Our distribution of
albedos formally is consistent with this prediction,
though it might require
many of our albedos
to be systematically too dark. A systematic error in 
albedo is indeed likely, due to the systematic errors
in $H$ (\S2.2.3). However, if the catalog $H$
values are systematically too bright then
the effect on our albedos would be to make them systematically
too bright (i.e., the opposite sense from making
our results align with the previous predictions).

There are a number of other results that arise from
our first 101~targets; here we present
a preliminary list of forthcoming papers
and results. (1) One of these 
objects, (85938) 1999~DJ4, is a known binary for which we can derive
the bulk density (Kistler et al., in prep.).
(2) We have observed roughly one dozen of our ground-truth
targets, and find that, overall, our results match
quite well to those obtained elsewhere (Harris et al., in prep.).
(3) Within our first 101~targets there are also a number of 
low delta~V targets -- that is, targets that spacecraft
can approach relatively easily (Mueller et al., in prep.).
Diameters for these objects are essential for
mission planning, and low albedos can indicate
primitive compositions, of interest for
various space missions (e.g., Marco Polo, Osiris-REx).
(4) We have begun a study of the albedos of objects in 
our sample with
low Tisserand parameter with respect to Jupiter,
with the idea of examining
whether, as predicted \citep{levison,binzel04}, these objects
preferentially derive from the outer Solar System (as suggested
by low albedos). In our first 101~targets, we have
only two targets with $T_J<3$ (the nominal
cutoff in \citet{levison,fernandez05,demeo08}), so our results will be presented
in a future paper (Harris et al., in prep.).
(5) Combining our Spitzer results with results
from our ground-based campaign (\S6) allows us to
evaluate the relationship between spectrally determined
compositions (through taxonomic typing) and
albedo and to measure the uncertainties
in the derived albedos for individual objects (Thomas et al., in
prep.)

\subsection{Results for individual objects}

\subsubsection{(433) Eros}

A great deal is known about (433) Eros from the NEAR spacecraft mission
and supporting ground-based observations. The interpretation of thermal-infrared
observations of (433) Eros is complicated by the fact that it can have a very
large lightcurve amplitude (up to 1.4 mag) and that its rotation axis lies near the
ecliptic plane
\citep[$\lambda$ = 17$^\circ$, $\beta$ = 11$^\circ$;][]{miller02}.
The mean diameter derived on the basis of the known volume
\citep{cheng02}
is 17.5~km, whereby previous thermal-infrared measurements
have given effective diameters of 14.3~km and 23.6~km at lightcurve minimum and
maximum, respectively \citep{harrisdavies}.
The disk-averaged albedo of Eros is around
0.22 \citep{harrisdavies,li04}.

The naive fixed-$\eta$ fit to our Eros data yields an albedo of~0.06 and a diameter
of 31.93~km, quite different from the values cited above. A detailed investigation
of these data and model results is therefore warranted, both to understand why
our model results do not reproduce the expected results, and more broadly to
characterize systematic uncertainties present in our modeling.

The line of sight from Spitzer to Eros was around 5~degrees different from Eros' rotation
axis at the time of the Warm Spitzer observations. With near pole-on geometry at a
phase angle of 36.5~degrees, the effects of rotation/thermal inertia on $\eta$ 
are much reduced since the thermal emission is concentrated in the visible hemisphere. Therefore, the adopted relation between $\eta$ and phase angle, which embodies a generalized first-order correction for such effects, is not applicable in this unusual case. \citet{harrisdavies} observed Eros at a geometry similar to that of our Spitzer observations (the line of sight to Eros was only 26~degrees away from the rotation axis, at a phase angle of 31~degrees), and found that $\eta = 1.07$ gave the best fit. Finally, the appropriate value for $H$ is not 11.16, the mean value employed above in our naive fit, but rather~10.46, which corresponds to lightcurve maximum, appropriate for a pole-on view. Using $\eta = 1.07$, $H = 10.46$, and the Spitzer 4.5~micron (thermal) flux, we find a (maximum) diameter of 22.3~km and \pv\ of~0.23, which are very close to the results of \citet{harrisdavies}
and exactly the albedo found by \citet{li04}.

A remaining outstanding issue is the 3.6~micron flux, which is larger than this model
would predict. One solution would be if the reflectance at 3.6~microns
is a
factor of two (or larger) greater than the reflectance at V~band; this
is significantly
different than our standard 40\% brighter assumption. Alternatively
the $G$~value (slope parameter),
which also influences the amount of reflected solar radiation,
may be different from the default value of $G = 0.15$ assumed
throughout this work.
We have found (Emery et al., in prep.) that
the 3.6~micron/V~reflectance ratio of Eros is about~1.65,
higher than~1.4 but not high enough to remove the problem of the
excess 3.6~micron flux.
However, taking a reflectance ratio of~1.65 we find that we can reproduce
the above values of
diameter, albedo, and $\eta$ for Eros by fitting NEATM to \emph{both} flux
measurements with
$G = 0.34$, a value that is much higher than our default value
of~0.15,
but still within the bounds of reasonable $G$~values, in light of the
value $G = 0.46$ given for
Eros by \citet{tedesco_mpc} (and note that
differences in viewing geometry can cause large variations in measured $G$~values).

Further investigation of the influence of assumed $G$~value on
floating-$\eta$ fits to our Warm
Spitzer data will be the subject of future work. Since the thermal
flux at 4.5~microns
is relatively insensitive to the 3.6~micron/V~reflectance ratio and
$G$,
the uncertainty introduced into the fixed-$\eta$ results by errors in
these parameters is very small.
For example, with $H = 10.46$, $G = 0.15$, and reflectance ratio
of~1.4,
the fixed-$\eta$ diameter result is 30.17~km.
Taking $H = 10.46$, $G = 0.34$, and reflectance ratio of~1.6,
the fixed-$\eta$ diameter becomes 30.01~km.

The larger conclusion of this analysis is that Eros should act
as a reminder of the effects of rotation vectors (as well as
cases where $G\neq0.15$) and therefore that
the results for individual objects should be treated with caution.
However, Eros is also a special case --- and especially
poorly fit --- in some regards. 
For example, Eros is unusually pole-on in our Spitzer observations.
The probability that a randomly oriented rotation
pole has an orientation angle less than or equal to $\theta$
is $L(\leq\theta) = (1-\cos\theta)$ \citep{tb06}. Therefore, if NEO rotation
axes are distributed isotropically, less than 1\% of all
NEOs should be observed within 5~degrees of their
rotation poles as Eros was. 
The actual probability is likely even less, since
YORP thermal torques will tend to move obliquity
values toward 0~or 180~degrees (that is, $\theta=90$~degrees)
\citep{vokrouhlicky03,bottke06}, and there
is some observational evidence of this
\citep{laspina,krysz}.
In other words, it is unlikely
that any other of our first 101~targets was observed
with this close a pole-on geometry.
Therefore,
our $\eta$/phase angle relationship is likely to be
generally appropriate, and
many of our other, naive fits are likely to be
much more acceptable. The case of (2100) Ra-Shalom is 
one of these.

\subsubsection{(2100) Ra-Shalom}

\begin{figure}
\includegraphics[angle=270,scale=0.5]{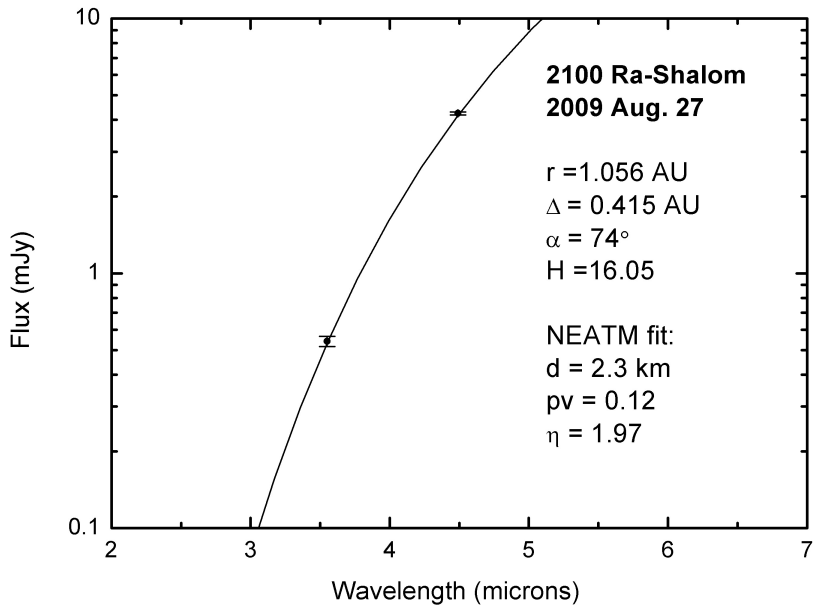}
\caption{NEATM fit to the Warm Spitzer thermal fluxes for (2100)~Ra-Shalom 
obtained in August, 2009.
(The reflected light component has been removed.)
This is one of the targets for which the quality of the data enables 
a floating-$\eta$ fit to be performed. 
The resulting diameter and albedo are in good agreement with
earlier groundbased results \citep{delbo03,shepard08}.
For most of our targets, especially those with high albedos,
the uncertainties in the 3.6~micron thermal flux components preclude 
a floating-$\eta$ NEATM fit, and we resort
to the fixed-$\eta$ method (\S4).
A comparison of the results obtained with these two methods, and results available in the
literature for some of our targets, will facilitate further refinement
of the fixed-$\eta$ method.
\label{rashalom}}
\end{figure}

(2100) Ra-Shalom is a well-studied object for which the diameter
and albedo have been established as a result of various infrared and
radar observing campaigns. It has a lightcurve amplitude of
0.41~mag
\citep{pravec98},
much lower than that of many other NEOs (compare to Eros, above, with
$\Delta$mag of~1.4).
(2100)~Ra-Shalom therefore serves as an excellent test of the viability of our
analysis procedures.
\citet{shepard08}
obtained an
effective diameter of 2.3$\pm$0.2~km and \pv\ of 0.13$\pm$0.03 from
radar observations.
\citet{harris98}
obtained virtually identical values from thermal-infrared
observations. The values obtained for (2100) Ra-Shalom from our warm
Spitzer data --- (2.35~km, \pv\ of~0.12) and (2.22~km, \pv\ of~0.14)
for floating-$\eta$ and fixed-$\eta$, respectively
(Table~\ref{comparison}) ---
agree remarkably well with the earlier results (Figure~\ref{rashalom}). 
Further work to check
the accuracy of our results against other ``ground-truth'' targets is underway
(Harris et al., in preparation).

\subsubsection{(5604) 1992 FE}

The $H$ value for (5604) 1992~FE provided by 
Horizons of~16.4 is probably incorrect.
\citet{delbo03}
used an updated value ($H=17.72$) obtained from the lightcurve observations of their
co-author P. Pravec. With this value they obtained an albedo of~0.48.
We use $H=17.72$ for the results presented in Table~\ref{comparison},
where we find a floating-$\eta$ solution with an albedo of~0.38.
This is more reasonable than the very large value (0.69)
given in Table~\ref{datatable}. This object is known
to be a V-type asteroid, so high albedos are expected.

\subsubsection{Objects with unusual albedos}

We derive \pv\ of~1.03 for (42286)~2001~TN41. This albedo
is almost certainly too large, as the largest plausible
albedo for an NEO is probably not too much larger than~0.5.
However, the total uncertainty in albedo for any given object
is around a factor of~2; at this level, the true albedo for 
2001~TN41 could be~0.5, which would be relatively
unsurprising. 
The CH1/CH2 flux ratio for this object is greater
than unity, which is unusual in our sample and
can only occur for
objects that are cold through being
distant or highly reflective
(see
Figure~\ref{sed}).
The large heliocentric distance of 2~AU, the
third largest within our sample, confirms
that this object is colder than most of our
other targets,
and its observed fluxes are
dominated by reflected sunlight (more than 90\% under our assumptions), rendering
the calculated thermal fluxes very sensitive to uncertainties in
$H$~magnitude.  The quantitative characterization of the resulting
statistical diameter uncertainty is beyond the scope of the current work
and will be treated in an upcoming paper (Mueller et al., in prep.).
When
more accurate $H$ is known, a more accurate albedo
can be derived.

The next highest derived albedos are 
0.8, 0.68, 0.6, and 0.49  for
(4953) 1990~MU, 2001~TX44,
(152637) 1997~NC1, and (10302) 1989~ML, respectively.
Aside from the first,
these albedos are generally plausible.
\citet{mueller07}
found \pv\ of~0.37 for 
(10302) 1989~ML; our derived value is some 30\% higher,
no larger than our expected albedo uncertainties.

Four objects in our first 101~targets have derived albedos
less than 2\%. These values are surprisingly low, and may
suffer from errors, especially in $H$ magnitude. However,
errors of 50\% or more, which would not be surprising
at this stage, would raise these albedos to 3\% or 4\%, values
that are consistent with expected values for dark
NEOs.

\subsubsection{Sub-kilometer NEOs}

Of our first 101~targets, more than half (56) have derived
diameters smaller than 1~km. The smallest object observed
to date is 2006~SY5, with a derived diameter of 87~meters
(with a very plausible albedo of~0.34).
With expected errors of perhaps 50\% on diameter, this
object is very likely to be smaller than 150~meters, and could
be as small as 50~meters, depending on the sign of potential
errors. These sub-kilometer bodies are among the smallest
NEOs for which physical properties are known. The distribution
of albedos for these 56~subkilometer objects is shown
in Figure~\ref{albhist}. The bias against small
objects with low albedos is readily apparent here.
Otherwise, these smallest NEOs seem to have
the same diverse compositions (as implied by a large
range of albedos) that the entire sample has.

\section{Future work}

The ExploreNEOs program has just begun.
Some analyses cannot be undertaken until
we have most or all of our data in hand.
These include a true analysis of the size
distribution of NEOs (a large sample size
is required for this); thermophysical modeling
of certain targets (detailed observations of
our ground-truth and multi-visit targets are needed);
determining the distribution of albedos in the
NEO population, though Figure~\ref{albhist} shows our
first results in this direction; and measuring
the mean albedo as a function of NEO size
(a large sample size is needed for this).

We have also begun a comprehensive ground-based
observing program that is a critical component
of the larger ExploreNEOs project. Because the
$H$ magnitudes of NEOs are typically quite poor,
we are carrying out a program of obtaining optical
photometry for (ideally) all of the targets in our 
program. We use a range of telescopes with apertures
from 0.36~meter to 8~meters. We use facilities to which
our team has institutional access as well as 
competed national telescopes and dedicated
facilities. As part of this
optical photometry program, we will obtain
lightcurves for a subset ($\sim$20\%) of our 
targets. We will use this lightcurve data set
as a proxy for the lightcurve variability inherent
in the larger sample, and thus to understand the
errors present in our model results. Both the optical
photometry catalog and the lightcurve database
will be interesting scientific contributions on their
own, of course, and will be presented in forthcoming
papers. At the conclusion of the ExploreNEOs program,
we will complete a final re-calculation of all model
results, incorporating results from this ground-based program.
Due to improved $H$ magnitudes,
those final errors should be substantially reduced from
those presented here.

Our ground-based campaign also includes obtaining
spectroscopy and spectrophotometry of several hundred
of our targets. The goal of these spectral measurements
is to determine taxa and compositions for those targets.
We will use these compositional determinations
together with our albedo results to understand the 
compositional distribution within the NEO population.
We will measure the correlation between spectrally-determined
compostions and albedos, and use this relationship
to determine by proxy the compositions for those objects
in our sample for which we will not obtain spectral
information, only albedo.

\section{Summary}

We have begun a large NEO survey called ExploreNEOs,
which uses Warm Spitzer to observe some 700~NEOs
at 3.6~and 4.5~microns.
We have also begun a comprehensive ground-based observing
program that complements the Warm Spitzer data that
we will obtain.
We have a large number of science
goals associated with ExploreNEOs; the overall goal
is to explore the history of near-Earth space.
We have presented here some fundamental
information about the design of our program.
We 
present in this paper results for the first 101~targets in our 
program. We find that the distribution of albedos
in this first sample is quite broad, probably 
indicating a wide range of compositions within the
NEO population.
Many objects smaller than one kilometer have
high albedos ($\gtrsim 0.35$), but
few objects larger than one kilometer
have high albedos. This result is consistent
with the idea that these objects are
collisionally older, and therefore possess surfaces
that are more space weathered and therefore darker,
or are not subject to other surface rejuvenating events
as frequently as smaller NEOs.
We find that our nominal results for (2100) Ra-Shalom
match previously published values quite closely,
while our nominal results for Eros diverge significantly
from previous values, though in a readily explainable way.
A number of objects have plausible but high ($>$0.5) 
albedos.
However, given the uncertainties present in our thermal
modeling at this time, all of these results are only preliminary.
We will continue to refine our thermal models and
improve the optical magnitudes for these targets
in order to minimize the uncertainties on diameter
and albedo.
This paper is the first in a series of papers that will
present various results from ExploreNEOs.

\acknowledgments

We acknowledge the thorough and prompt hard work of the
staff at the Spitzer Science Center, without whom
the execution of this program would not be possible.
We would like to thank Michael Mommert (DLR) for help in checking the modeling results given in the data tables.
We thank an anonymous referee for making a number
of useful suggestions.
This work is based in part on observations made with the Spitzer Space Telescope, which is operated by JPL/Caltech under a contract with NASA. Support for this work was provided by NASA through an award issued by JPL/Caltech.

Facilities: \facility{Spitzer(IRAC)}

\clearpage



\clearpage


\clearpage

\begin{deluxetable}{llllcccccccccl}
\tabletypesize{\tiny}
\rotate
\tablecaption{Target information and results \label{datatable}}
\tablewidth{0pt}
\tablehead{
\colhead{Target} & \colhead{PID} & \colhead{AOR} & \colhead{Date obs.} & \colhead{F3.6} &
\colhead{F4.5} & \colhead{r} & \colhead{$\Delta$} &
\colhead{phase} & \colhead{\pv } &
\colhead{$D$} & \colhead{$\eta$} & \colhead{$H$} & \colhead{Notes} \\
\colhead{} & \colhead{} & \colhead{} & \colhead{(UT)} & \colhead{($\mu$Jy)} &
\colhead{($\mu$Jy)} & \colhead{(AU)} & \colhead{(AU)} &
\colhead{(deg)} & \colhead{} &
\colhead{(km)} & \colhead{} & \colhead{(mag)} & \colhead{}
}
\startdata
433 Eros 1898 DQ & 60012 & 32648448 & 2009-Aug-25T06:16:48 & 13990$\pm$110 & 40230$\pm$200 & 1.702 & 1.218 & 36.50 & 0.06 & 31.93 & 1.39 & 11.2 & 1 \\ 
1863 Antinous 1948 EA & 60012 & 32698112 & 2009-Sep-02T12:50:32 & 5422$\pm$71 & 21070$\pm$130 & 1.024 & 0.240 & 83.28 & 0.10 & 3.23 & 1.99 & 15.5 \\ 
1943 Anteros 1973 EC & 60012 & 32641536 & 2009-Sep-15T00:09:48 & 189$\pm$13 & 582$\pm$22 & 1.548 & 0.951 & 40.17 & 0.15 & 2.43 & 1.43 & 15.8 \\ 
2100 Ra-Shalom 1978 RA & 61012 & 35304192 & 2009-Aug-27T01:22:49 & 937$\pm$29 & 4841$\pm$63 & 1.056 & 0.415 & 74.12 & 0.14 & 2.22 & 1.87 & 16.1 & 1\\ 
3103 Eger 1982 BB & 60012 & 32584448 & 2009-Aug-24T22:47:37 & 162$\pm$12 & 243$\pm$14 & 1.584 & 1.011 & 39.14 & 0.39 & 1.79 & 1.42 & 15.4 \\ 
4183 Cuno 1959 LM & 60012 & 32591104 & 2009-Aug-20T09:41:02 & 4387$\pm$61 & 18420$\pm$130 & 1.171 & 0.524 & 60.71 & 0.10 & 5.49 & 1.70 & 14.4 \\ 
4953 1990 MU & 60012 & 32654336 & 2009-Aug-21T00:58:18 & 16.3$\pm$3.9 & 17.9$\pm$4.0 & 2.667 & 2.091 & 20.61 & 0.79 & 2.26 & 1.18 & 14.1  \\ 
4957 Brucemurray 1990 XJ & 60012 & 32616448 & 2009-Aug-13T04:04:51 & 81.6$\pm$8.7 & 193$\pm$13 & 1.781 & 1.589 & 34.69 & 0.17 & 3.06 & 1.36 & 15.1  \\ 
5143 Heracles 1991 VL & 60012 & 32667648 & 2009-Sep-19T01:20:11 & 2211$\pm$44 & 8883$\pm$89 & 1.080 & 0.435 & 71.12 & 0.42 & 3.26 & 1.83 & 14.0  \\ 
5604 1992 FE & 60012 & 32598272 & 2009-Aug-06T13:52:59 & 510$\pm$21 & 1868$\pm$39 & 1.061 & 0.210 & 73.26 & 0.69 & 0.84 & 1.86 & 16.4 & 1 \\ 
5620 Jasonwheeler 1990 OA & 60012 & 32659968 & 2009-Aug-25T07:00:35 & 258$\pm$16 & 1252$\pm$33 & 1.321 & 0.620 & 48.68 & 0.09 & 1.75 & 1.54 & 17.0  \\ 
5626 1991 FE & 60012 & 32574464 & 2009-Aug-21T03:23:36 & 434$\pm$20 & 1209$\pm$31 & 1.689 & 0.972 & 33.08 & 0.15 & 3.96 & 1.34 & 14.7  \\  
7822 1991 CS & 60012 & 32693248 & 2009-Aug-25T04:37:42 & 141$\pm$13 & 420$\pm$19 & 1.122 & 0.514 & 65.61 & 0.28 & 0.83 & 1.76 & 17.4  \\ 
7888 1993 UC & 60012 & 32576768 & 2009-Aug-06T17:17:59 & 142$\pm$12 & 456$\pm$20 & 1.535 & 1.213 & 41.56 & 0.18 & 2.75 & 1.45 & 15.3  \\ 
8566 1996 EN & 60012 & 32651264 & 2009-Oct-17T05:43:42 & 80.3$\pm$8.7 & 223$\pm$14 & 1.317 & 0.963 & 50.63 & 0.28 & 1.25 & 1.57 & 16.5  \\ 
9950 ESA 1990 VB & 60012 & 32644352 & 2009-Oct-06T01:02:44 & 2386$\pm$45 & 9875$\pm$89 & 1.216 & 0.360 & 50.59 & 0.10 & 2.47 & 1.57 & 16.2  \\ 
10302 1989 ML & 60012 & 32642048 & 2009-Aug-21T03:09:36 & 229$\pm$15 & 560$\pm$22 & 1.100 & 0.152 & 56.01 & 0.49 & 0.24 & 1.64 & 19.5  \\ 
12711 1991 BB & 60012 & 32593664 & 2009-Oct-16T16:45:35 & 371$\pm$19 & 1242$\pm$32 & 1.172 & 0.708 & 60.28 & 0.19 & 1.94 & 1.69 & 16.0  \\ 
15745 1991 PM5 & 60012 & 32624384 & 2009-Aug-25T06:47:16 & 66.2$\pm$9.3 & 209$\pm$14 & 1.361 & 0.632 & 45.54 & 0.23 & 0.77 & 1.50 & 17.8  \\ 
16834 1997 WU22 & 60012 & 32666112 & 2009-Sep-06T17:43:13 & 67.1$\pm$9.3 & 134$\pm$12 & 1.514 & 1.225 & 42.30 & 0.40 & 1.52 & 1.46 & 15.7  \\ 
17274 2000 LC16 & 60012 & 32685824 & 2009-Aug-25T06:31:32 & 1711$\pm$39 & 7533$\pm$83 & 1.201 & 0.753 & 57.88 & 0.01 & 5.01 & 1.66 & 16.7  \\
20460 Robwhiteley 1999 LO28 & 60012 & 32597504 & 2009-Aug-20T08:56:45 & 2011$\pm$42 & 8146$\pm$82 & 1.239 & 0.634 & 55.40 & 0.04 & 4.57 & 1.63 & 15.7  \\ 
20826 2000 UV13 & 60012 & 32634880 & 2009-Aug-20T23:32:41 & 33.3$\pm$5.5 & 31.8$\pm$5.3 & 3.190 & 2.640 & 17.09 & 0.27 & 5.12 & 1.13 & 13.5  \\ 
27346 2000 DN8 & 60012 & 32587008 & 2009-Aug-22T21:19:03& 142$\pm$12 & 485$\pm$21 & 1.350 & 0.997 & 48.88 & 0.19 & 1.92 & 1.54 & 16.0  \\ 
38086 Beowolf 1999 JB & IWIC & IWIC & 2009-Jul-26T00:27:03 & 26.4$\pm$5.3 & 156.7$\pm$4.4 & 1.187 & 0.715 & 58.64 & 0.37 & 0.72 & 1.67 & 17.4  \\ 
40329 1999 ML & 60012 & 32603648 & 2009-Sep-03T00:25:52 & 197$\pm$14 & 623$\pm$23 & 1.344 & 0.501 & 41.48 & 0.16 & 0.94 & 1.45 & 17.7  \\ 
42286 2001 TN41 & 61013 & 32717568 & 2009-Oct-26T23:05:59 & 24.7$\pm$4.7 & 19.3$\pm$4.2 & 1.961 & 1.236 & 26.92 & 1.03 & 0.69 & 1.26 & 16.4  & 1\\ 
52387 1993 OM7 & 60012 & 32558592 & 2009-Sep-01T10:12:03 & 95.1$\pm$9.6 & 333$\pm$17 & 1.520 & 0.727 & 35.88 & 0.09 & 1.22 & 1.38 & 17.8  \\ 
54401 2000 LM & 60012 & 32578048 & 2009-Oct-26T22:05:42 & 50.2$\pm$7.0 & 118$\pm$10 & 1.494 & 0.890 & 42.10 & 0.22 & 0.98 & 1.46 & 17.3  \\ 
54686 2001 DU8 & 60012 & 32586496 & 2009-Oct-29T01:20:33 & 65.8$\pm$8.2 & 96.6$\pm$9.3 & 1.749 & 1.019 & 31.29 & 0.34 & 1.25 & 1.32 & 16.3 & 2 \\ 
55532 2001 WG2 & 60012 & 32642816 & 2009-Aug-28T01:26:05 & 285$\pm$17 & 985$\pm$30 & 1.223 & 0.792 & 56.28 & 0.14 & 1.96 & 1.64 & 16.3  \\ 
65679 1989 UQ & 60012 & 32638976 & 2009-Oct-15T20:50:49 & 405$\pm$61 & 2264$\pm$43 & 1.129 & 0.237 & 58.64 & 0.06 & 0.71 & 1.67 & 19.4  \\ 
67381 2000 OL8 & 60012 & 32635648 & 2009-Aug-25T05:03:54 & 22.6$\pm$4.5 & 90.8$\pm$8.7 & 1.134 & 0.381 & 63.58 & 0.25 & 0.28 & 1.74 & 19.9  \\ 
68372 2001 PM9 & 60012 & 32563968 & 2009-Aug-20T10:32:54& 3928$\pm$59 & 19610$\pm$140 & 1.130 & 0.204 & 53.62 & 0.02 & 1.70 & 1.61 & 18.9  \\ 
85938 1999 DJ4 & 61013 & 32720896 & 2009-Aug-13T15:40:18 & 17.2$\pm$3.9 & 62.0$\pm$7.2 & 1.409 & 0.688 & 43.10 & 0.28 & 0.48 & 1.47 & 18.6  \\ 
85989 1999 JD6 & 60012 & 32661248 & 2009-Aug-31T12:59:55 & 1618$\pm$37 & 8506$\pm$85 & 1.261 & 0.390 & 45.22 & 0.04 & 2.39 & 1.50 & 17.1  \\ 
90367 2003 LC5 & 60012 & 32583424 & 2009-Aug-24T23:16:30 & 279$\pm$16 & 1590$\pm$37 & 1.391 & 0.705 & 45.06 & 0.03 & 2.39 & 1.50 & 17.7  \\ 
99799 2002 LJ3 & 60012 & 32613120 & 2009-Sep-19T22:38:45 & 125$\pm$15 & 392$\pm$20 & 1.238 & 0.354 & 46.18 & 0.42 & 0.49 & 1.51 & 18.1  \\ 
108519 2001 LF & 60012 & 32636160 & 2009-Aug-13T13:09:08 & 1707$\pm$39 & 9686$\pm$87 & 1.240 & 0.366 & 46.19 & 0.02 & 2.31 & 1.51 & 17.9  \\ 
138883 2000 YL29 & 60012 & 32592640 & 2009-Oct-15T22:49:18 & 108$\pm$10 & 318$\pm$17 & 1.319 & 0.902 & 50.91 & 0.19 & 1.39 & 1.57 & 16.7  \\ 
138911 2001 AE2 & 60012 & 32662272 & 2009-Aug-13T14:53:37 & 24.4$\pm$4.7 & 88.8$\pm$8.6 & 1.330 & 0.483 & 41.78 & 0.34 & 0.34 & 1.45 & 19.1  \\ 
138925 2001 AU43 & 60012 & 32616960 & 2009-Aug-25T03:00:42 & 535$\pm$23 & 2389$\pm$45 & 1.182 & 0.645 & 59.84 & 0.11 & 2.41 & 1.69 & 16.1  \\ 
138937 2001 BK16 & 60012 & 32672256 & 2009-Aug-23T12:50:06 & 87.4$\pm$8.9 & 359$\pm$17 & 1.182 & 0.629 & 59.86 & 0.21 & 0.92 & 1.69 & 17.5  \\ 
143651 2003 QO104 & 60012 & 32694272 & 2009-Aug-21T02:40:18 & 276$\pm$42 & 1052$\pm$31 & 1.402 & 0.801 & 45.93 & 0.13 & 2.31 & 1.51 & 16.0  \\ 
152637 1997 NC1 & 60012 & 32675584 & 2009-Aug-22T20:20:23 & 1344$\pm$35 & 3434$\pm$55 & 1.046 & 0.088 & 72.53 & 0.59 & 0.43 & 1.85 & 18.0  \\ 
153220 2000 YN29 & IWIC & IWIC & 2009-Jul-26T00:03:10 & 2877.1$\pm$8.6 & 10928$\pm$11 & 1.068 & 0.119 & 62.27 & 0.27 & 0.81 & 1.72 & 17.5  \\ 
159399 1998 UL1 & 60012 & 32701184 & 2009-Aug-13T06:56:16 & 46.7$\pm$8.2 & 110$\pm$10 & 1.502 & 1.130 & 42.84 & 0.27 & 1.24 & 1.47 & 16.6  \\ 
159402 1999 AP10 & 60012 & 32682752 & 2009-Aug-31T22:54:56 & 90$\pm$10 & 278$\pm$16 & 1.292 & 0.838 & 52.32 & 0.34 & 1.20 & 1.59 & 16.4  \\ 
159608 2002 AC2 & 60012 & 32590848 & 2009-Aug-25T00:11:43 & 6242$\pm$75 & 27020$\pm$160 & 1.108 & 0.157 & 53.63 & 0.20 & 1.51 & 1.61 & 16.5  \\ 
162195 1999 RK45 & 60012 & 32689408 & 2009-Nov-03T14:22:00 & 476$\pm$21 & 1602$\pm$37 & 1.088 & 0.147 & 60.85 & 0.19 & 0.39 & 1.70 & 19.5  \\ 
162483 2000 PJ5 & 60012 & 32579840 & 2009-Aug-21T02:55:14 & 287$\pm$17 & 1026$\pm$30 & 1.196 & 0.393 & 55.03 & 0.21 & 0.92 & 1.62 & 17.5  \\ 
163758 2003 OS13 & 60012 & 32601856 & 2009-Jul-30T15:39:01 & 17.5$\pm$3.9 & 45.3$\pm$6.1 & 1.538 & 0.922 & 39.88 & 0.44 & 0.66 & 1.43 & 17.4  \\ 
164201 2004 EC & 60012 & 32613888 & 2009-Jul-30T18:51:35 & 845$\pm$27 & 5092$\pm$66 & 1.158 & 0.560 & 61.67 & 0.12 & 2.87 & 1.71 & 15.6  \\ 
164202 2004 EW & 60012 & 32667392 & 2009-Aug-06T05:32:19 & 45.9$\pm$6.2 & 135$\pm$10 & 1.049 & 0.158 & 75.37 & 0.35 & 0.16 & 1.89 & 20.8  \\ 
177614 2004 HK33 & 60012 & 32569600 & 2009-Jul-30T16:25:59 & 5895$\pm$71 & 22920$\pm$140 & 1.057 & 0.095 & 63.88 & 0.19 & 0.93 & 1.74 & 17.6  \\ 
184990 2006 KE89 & 60012 & 32604672 & 2009-Aug-14T08:13:19 & 103$\pm$11 & 371$\pm$18 & 1.228 & 0.805 & 55.75 & 0.29 & 1.24 & 1.64 & 16.5  \\ 
1996 EO & 61013 & 32724992 & 2009-Sep-06T18:56:13 & 25.4$\pm$4.7 & 99.3$\pm$9.1 & 1.151 & 0.540 & 62.84 & 0.24 & 0.42 & 1.73 & 19.1  \\ 
1996 TY11 & 60012 & 32576256 & 2009-Sep-01T19:41:57 & 230$\pm$14 & 941$\pm$28 & 1.056 & 0.248 & 75.79 & 0.09 & 0.62 & 1.90 & 19.3  \\ 
1998 QE2 & 60012 & 32680960 & 2009-Sep-23T02:32:45 & 514$\pm$22 & 2562$\pm$46 & 1.183 & 0.728 & 59.35 & 0.06 & 2.75 & 1.68 & 16.4  \\ 
1998 SV4 & 60012 & 32625664 & 2009-Oct-06T11:24:45 & 79.3$\pm$8.3 & 327$\pm$16 & 1.321 & 0.530 & 46.02 & 0.20 & 0.74 & 1.51 & 18.0  \\ 
1999 TX2 & 60012 & 32564736 & 2009-Oct-26T21:29:07 & 326$\pm$18 & 1299$\pm$34 & 1.142 & 0.362 & 62.24 & 0.11 & 0.96 & 1.72 & 18.1  \\ 
2000 GV147 & 60012 & 32620288 & 2009-Oct-24T06:09:18 & 100$\pm$10 & 337$\pm$17 & 1.064 & 0.336 & 74.34 & 0.17 & 0.50 & 1.88 & 19.0  \\ 
2000 JY8 & 60012 & 32700416 & 2009-Jul-29T14:18:43 & 23.7$\pm$4.8 & 100.4$\pm$9.1 & 1.555 & 1.018 & 40.26 & 0.32 & 1.11 & 1.43 & 16.6  \\ 
2001 TX44 & 60012 & 32693760 & 2009-Oct-27T00:09:19 & 13.9$\pm$3.5 & 37.3$\pm$5.6 & 1.332 & 0.499 & 43.09 & 0.68 & 0.26 & 1.47 & 19.0  \\ 
2002 EZ2 & 60012 & 32648960 & 2009-Aug-24T11:00:05 & 14.0$\pm$3.6 & 63.7$\pm$7.3 & 1.241 & 0.367 & 46.42 & 0.40 & 0.21 & 1.51 & 20.0  \\ 
2002 FB6 & 60012 & 32559360 & 2009-Oct-06T12:23:39 & 264$\pm$16 & 10593$\pm$30 & 1.078 & 0.477 & 70.85 & 0.14 & 1.20 & 1.83 & 17.4  \\ 
2002 GO5 & 60012 & 32589056 & 2009-Oct-16T12:13:04 & 278$\pm$17 & 941$\pm$28 & 1.218 & 0.351 & 49.66 & 0.24 & 0.75 & 1.56 & 17.8  \\ 
2002 HF8 & 60012 & 32562432 & 2009-Aug-24T23:44:38 & 109.3$\pm$9.9 & 455$\pm$20 & 1.262 & 0.449 & 48.69 & 0.18 & 0.70 & 1.54 & 18.3  \\ 
2002 LS32 & 60012 & 32658688 & 2009-Aug-24T12:14:18 & 720$\pm$26 & 2993$\pm$51 & 1.098 & 0.163 & 58.72 & 0.29 & 0.57 & 1.67 & 18.2  \\ 
2002 LV & 60012 & 32618240 & 2009-Aug-06T05:01:48 & 456$\pm$20 & 1825$\pm$38 & 1.101 & 0.537 & 67.11 & 0.15 & 1.73 & 1.78 & 16.5  \\ 
2002 MQ3 & 60012 & 32682496 & 2009-Aug-22T20:35:42 & 176$\pm$13 & 685$\pm$24 & 1.114 & 0.290 & 64.52 & 0.32 & 0.57 & 1.75 & 18.1  \\ 
2002 NP1 & 60012 & 32556800 & 2009-Nov-04T00:16:49 & 137$\pm$12 & 474$\pm$20 & 1.190 & 0.479 & 58.47 & 0.25 & 0.81 & 1.67 & 17.6  \\ 
2002 OM4 & 60012 & 32668928 & 2009-Sep-19T01:41:07 & 67.2$\pm$7.8 & 256$\pm$15 & 1.274 & 0.772 & 53.56 & 0.30 & 1.03 & 1.61 & 16.9  \\ 
2002 QE7 & 60012 & 32637952 & 2009-Sep-23T04:42:22 & 66.2$\pm$8.0 & 188$\pm$13 & 1.217 & 0.329 & 47.87 & 0.34 & 0.31 & 1.53 & 19.3  \\ 
2002 TE66 & 60012 & 32696576 & 2009-Oct-30T05:22:11 & 23.6$\pm$4.8 & 69.5$\pm$7.7 & 1.275 & 0.668 & 53.18 & 0.39 & 0.48 & 1.60 & 18.2  \\ 
2002 UX & 60012 & 32571136 & 2009-Aug-25T03:17:58 & 157$\pm$13 & 443$\pm$20 & 1.285 & 0.422 & 43.82 & 0.31 & 0.65 & 1.48 & 17.8  \\ 
2003 SL5 & 60012 & 32626944 & 2009-Sep-19T23:06:20 & 289$\pm$17 & 1052$\pm$30 & 1.114 & 0.164 & 53.62 & 0.36 & 0.33 & 1.61 & 19.1  \\ 
2003 TJ2 & 60012 & 32632064 & 2009-Oct-29T23:14:50 & 46.5$\pm$6.6 & 226$\pm$14 & 1.048 & 0.369 & 76.19 & 0.23 & 0.45 & 1.90 & 18.9  \\ 
2003 TL4 & 60012 & 32567552 & 2009-Oct-16T16:59:20 & 131$\pm$12 & 544$\pm$22 & 1.042 & 0.185 & 79.79 & 0.22 & 0.38 & 1.95 & 19.4  \\ 
2003 UN12 & 60012 & 32628736 & 2009-Oct-06T22:51:40 & 257$\pm$15 & 1506$\pm$35 & 1.302 & 0.461 & 44.48 & 0.05 & 1.25 & 1.49 & 18.5  \\ 
2003 WD158 & 60012 & 32601088 & 2009-Jul-31T00:06:53 & 101$\pm$11 & 402$\pm$19 & 1.170 & 0.316 & 54.42 & 0.29 & 0.44 & 1.62 & 18.8 & 3 \\ 
2003 WD158 & IWIC & IWIC & 2009-Jul-25T22:47:12 & 141.0$\pm$5.9 & 246.0$\pm$4.9 & 1.138 & 0.312 & 60.06 & 0.39 & 0.38 & 1.69 & 18.8 & 3 \\ 
2003 WO7 & 60012 & 32568576 & 2009-Oct-09T15:38:14 & 225$\pm$22 & 517$\pm$24 & 1.237 & 0.418 & 50.83 & 0.10 & 0.68 & 1.57 & 18.9  \\ 
2004 FU64 & 60012 & 32557056 & 2009-Aug-22T20:05:13 & 175$\pm$14 & 554$\pm$22 & 1.169 & 0.499 & 60.79 & 0.19 & 0.91 & 1.70 & 17.6  \\ 
2004 JN13 & 60012 & 32667904 & 2009-Aug-28T21:14:08 & 429$\pm$20 & 1608$\pm$37 & 1.559 & 0.774 & 34.76 & 0.25 & 2.97 & 1.36 & 14.5  \\ 
2004 JX20 & 60012 & 32557824 & 2009-Jul-29T13:52:44 & 855$\pm$27 & 4468$\pm$58 & 1.116 & 0.314 & 64.11 & 0.01 & 1.48 & 1.74 & 19.3 & 3 \\ 
2004 JX20 & IWIC & IWIC & 2009-Jul-25T23:17:44 & 1181.8$\pm$2.4 & 3374.0$\pm$3.4 & 1.110 & 0.301 & 64.77 & 0.02 & 1.30 & 1.75 & 19.3 & 3 \\ 
2004 SB20 & 60012 & 32603904 & 2009-Oct-16T09:41:45 & 47.3$\pm$7.0 & 136$\pm$11 & 1.134 & 0.454 & 64.66 & 0.41 & 0.43 & 1.75 & 18.4  \\ 
2005 LY19 & 61013 & 32716288 & 2009-Oct-06T21:07:34 & 33.4$\pm$4.1 & 54.9$\pm$6.8 & 1.944 & 1.217 & 27.22 & 0.26 & 1.36 & 1.26 & 16.4  \\ 
2005 OU2 & 61013 & 32724224 & 2009-Oct-03T05:24:48 & 43.2$\pm$6.1 & 124$\pm$10 & 1.172 & 0.394 & 58.94 & 0.37 & 0.34 & 1.68 & 19.0  \\ 
2005 UH6 & 60012 & 32593920 & 2009-Oct-15T08:18:15 & 134$\pm$12 & 608$\pm$23 & 1.079 & 0.267 & 71.43 & 0.22 & 0.52 & 1.84 & 18.7  \\ 
2005 WC1 & 60012 & 32642560 & 2009-Aug-01T06:00:54 & 58.0$\pm$6.9 & 255$\pm$14 & 1.099 & 0.259 & 65.49 & 0.11 & 0.29 & 1.76 & 20.7  \\ 
2006 AD & 60012 & 32583168 & 2009-Sep-09T14:48:28 & 1500$\pm$36 & 6168$\pm$74 & 1.092 & 0.511 & 68.81 & 0.04 & 3.06 & 1.80 & 16.7  \\ 
2006 LF & 60012 & 32610304 & 2009-Jul-30T16:10:39 & 161$\pm$13 & 711$\pm$24 & 1.195 & 0.299 & 47.90 & 0.25 & 0.50 & 1.53 & 18.6 & 3 \\ 
2006 LF & IWIC & IWIC & 2009-Jul-25T23:46:15 & 233.9$\pm$9.0 & 899.8$\pm$7.2 & 1.148 & 0.278 & 55.93 & 0.20 & 0.56 & 1.64 & 18.6 & 3 \\ 
2006 MD12 & 60012 & 32600576 & 2009-Sep-01T09:43:13 & 36.4$\pm$5.9 & 117$\pm$10 & 1.155 & 0.332 & 58.91 & 0.43 & 0.27 & 1.68 & 19.4  \\ 
2006 NL & 60012 & 32694528 & 2009-Aug-13T14:09:58 & 24.6$\pm$5.0 & 144$\pm$11 & 1.106 & 0.242 & 63.63 & 0.46 & 0.22 & 1.74 & 19.7  \\ 
2006 OD7 & 60012 & 32628224 & 2009-Sep-19T12:14:38 & 195$\pm$13 & 615$\pm$23 & 1.113 & 0.197 & 58.96 & 0.27 & 0.33 & 1.68 & 19.5  \\ 
2006 SV19 & 60012 & 32702976 & 2009-Sep-15T16:08:48 & 109$\pm$10 & 435$\pm$19 & 1.166 & 0.672 & 60.97 & 0.12 & 1.06 & 1.70 & 17.8  \\ 
2006 SY5 & 60012 & 32647168 & 2009-Jul-30T18:18:27 & 31.2$\pm$5.1 & 116.7$\pm$9.7 & 1.093 & 0.136 & 54.07 & 0.34 & 0.09 & 1.61 & 22.1 & 1 \\ 
2006 WO127 & 60012 & 32638208 & 2009-Nov-04T01:09:24 & 103$\pm$11 & 350$\pm$18 & 1.528 & 0.889 & 40.23 & 0.21 & 1.69 & 1.43 & 16.2  \\ 
2007 DK8 & 60012 & 32575488 & 2009-Sep-16T06:56:22 & 322$\pm$17 & 1708$\pm$38 & 1.173 & 0.298 & 53.68 & 0.08 & 0.79 & 1.61 & 18.9  \\ 
2007 VD12 & 60012 & 32687104 & 2009-Jul-28T14:02:11 & 30.0$\pm$5.0 & 78.7$\pm$8.0 & 1.232 & 0.351 & 45.88 & 0.39 & 0.22 & 1.51 & 20.0  \\ 
2007 YQ56 & 60012 & 32639488 & 2009-Aug-20T10:04:58 & 79.5$\pm$8.6 & 353$\pm$17 & 1.029 & 0.199 & 82.47 & 0.16 & 0.34 & 1.98 & 20.0  \\ 
\enddata


\tablecomments{The columns are as follows: target name;
Spitzer program ID; AOR, which is a unique Spitzer identifier for
each observation; date and time observed, which is accurate
to a few seconds at 3.6~microns and tens of seconds
at 4.5~microns, and which is not corrected
for light travel time; measured fluxes at 3.6~and 
4.5~microns; heliocentric and Spitzer-centric distances; phase
angle of observations; derived albedo and diameter, using the
$\eta$ value given; absolute magnitude ($H$), from Horizons,
reported only to a single decimal place;
and additional notes (if any). 
The $\eta$ values given here 
represent the fixed-$\eta$ solutions and are derived
from the phase angle (\S5.1).
Uncertainties on individual diameter and albedo values
are estimated to be 25\% and 50\%, respectively.
Notes:
(1) This object is discussed in some detail in Section~5.3.
(2) CH1 possibly affected by background features.
(3) Target observed twice: once during IWIC and once during
nominal post-cryo operations.}
\end{deluxetable}

\clearpage

\begin{deluxetable}{lcccccc}
\tablecaption{Comparison between floating-$\eta$ and fixed-$\eta$ model results \label{comparison}}
\tablewidth{0pt}
\tablehead{
\colhead{} & \multicolumn{3}{c}{NEATM floating-$\eta$} & \multicolumn{3}{c}{Fixed-$\eta$} \\
\colhead{Target} & \colhead{$D$} & \colhead{\pv} & \colhead{$\eta$} & \colhead{$D$} &
\colhead{\pv} & \colhead{$\eta$} \\
\colhead{} & \colhead{(km)} & \colhead{} & \colhead{} & \colhead{(km)} &
\colhead{} & \colhead{}}
\startdata
2100 Ra-Shalom 1978 RA & 2.35 & 0.12 & 1.97 & 2.22 & 0.14 & 1.87 \\
5604 1992 FE\tablenotemark{a} & 0.61 & 0.38 & 1.56 & 0.71 & 0.28 & 1.86 \\
67381 2000 OL8 & 0.32 & 0.19 & 1.98 & 0.28 & 0.25 & 1.74 \\
138937 2001 BK16 & 0.83 & 0.26 & 1.54 & 0.92 & 0.21 & 1.69 \\
164201 2004 EC & 4.34 & 0.05 & 2.35 & 2.87 & 0.12 & 1.71 \\
184990 2006 KE89 & 1.59 & 0.18 & 2.06 & 1.24 & 0.29 & 1.64 \\
1996 EO & 0.39 & 0.27 & 1.63 & 0.42 & 0.24 & 1.73 \\
1998 SV4 & 0.71 & 0.21 & 1.46 & 0.74 & 0.20 & 1.51 \\
2002 HF8 & 0.55 & 0.29 & 1.23 & 0.70 & 0.18 & 1.54 \\
2002 LS32 & 0.82 & 0.14 & 2.33 & 0.57 & 0.29 & 1.67 \\
2003 SL5 & 0.50 & 0.16 & 2.41 & 0.33 & 0.36 & 1.61 \\
2003 TL4 & 0.37 & 0.24 & 1.89 & 0.38 & 0.22 & 1.95 \\
2003 UN12 & 0.98 & 0.08 & 1.24 & 1.25 & 0.05 & 1.49 \\
2003 WD158 & 0.67 & 0.12 & 2.38 & 0.44 & 0.29 & 1.62 \\
2004 SB20 & 0.49 & 0.32 & 2.00 & 0.43 & 0.41 & 1.75 \\
2005 OU2 & 0.26 & 0.63 & 1.22 & 0.34 & 0.37 & 1.68 \\
2005 UH6 & 0.70 & 0.12 & 2.41 & 0.52 & 0.22 & 1.84 \\
2006 LF & 0.77 & 0.11 & 2.22 & 0.50 & 0.25 & 1.53 \\
2006 SY5 & 0.12 & 0.19 & 2.15 & 0.09 & 0.34 & 1.61 \\
\enddata
\tablenotetext{a}{Note: We use $H=17.72$ for (5604) 1992~FE (see Section~5.3).}
\tablecomments{The mean fractional difference between the two sets
of diameters, $((D_{fixed}-D_{float})/D_{float})$, is -7.5\%;
the mean absolute fractional difference,
$(|(D_{fixed}-D_{float})|/D_{float})$, is 21\%.}
\end{deluxetable}





\end{document}